\documentclass[a4paper,fleqn,usenatbib]{mnras}

\usepackage[T1]{fontenc}
\usepackage{ae,aecompl}

\usepackage{graphicx}	
\usepackage{amsmath}	
\usepackage{amssymb}	
\usepackage{txfonts}
\usepackage{amsfonts}
\usepackage{amsbsy}
\usepackage{amsmath}
\usepackage{pdflscape}

\newcommand{\bx}{\bar{X}}
\newcommand{\bbsigma}{\bar{\sigma}}

\newcommand{\bbchi}{\mbox{\Large$\chi$}}

\newcommand{\nb}{\boldsymbol{n}}
\newcommand{\ssb}{\boldsymbol{s}}

\newcommand{\xb}{\boldsymbol{x}}

\newcommand{\Nc}{\mathcal{N}}
\newcommand{\Mc}{\mathcal{M}}
\newcommand{\bbf}{\mathfrak{b}}
\newcommand{\gf}{\mathfrak{g}}
\newcommand{\ff}{\mathfrak{f}}
\newcommand{\Zf}{\mathrm{Z}}
\newcommand{\zf}{\mathrm{z}}

\title[Spectral analysis of uneven time series]{A critical comparison of the Lomb-Scargle and the classical periodograms}

\author[R. Vio \& P. Andreani]{
Roberto Vio,$^{1}$\thanks{robertovio@tin.it}
P. Andreani,$^{2}$
\\
$^{1}$Chip Computers Consulting s.r.l., Viale Don L.~Sturzo 82, S.Liberale di Marcon, 30020 Venice, Italy\\
$^{2}$ESO, Karl Schwarzschild strasse 2, 85748 Garching, Germany
}

\date{Accepted XXX. Received YYY; in original form ZZZ}

\pubyear{2020}

\begin{document}
\label{firstpage}
\pagerange{\pageref{firstpage}--\pageref{lastpage}}
\maketitle

\begin{abstract}
The detection of signals hidden in noise is one of the oldest and common problems in astronomy. Various solutions  have been proposed in the past such as the parametric approaches based on the least-squares fit 
of theoretical templates or the non-parametric techniques as the phase-folding method. Most of them, however, are suited only for signals with specific time evolution. For generic signals the spectral approach based on the periodogram is potentially the most effective. In astronomy the main problem in working with the periodogram is that often the sampling of the signals is irregular. This complicates its efficient computation (the fast Fourier transform cannot be directly used) but overall the determination of its statistical characteristics. The Lomb-Scargle periodogram (LSP) provides a solution to this last important issue, but its main drawback is the assumption of a very specific model of the data
which is not correct for most of the practical applications. These issues are not always considered in literature with theoretical and practical consequences of no easy solution. 
Moreover, apart from pathological samplings, it is common believe that the LSP and the classical periodogram (CP) usually provide almost identical results. In general, this is true but here it is shown that there are situations where the LSP is less effective than the CP in the detection of signals in noise. There are no compelling reasons, therefore, to use the LSP instead of the CP which is directly connected to the correlation function of the observed signal with the sinusoidal functions at the various frequencies of interest.
\end{abstract}




\section{Introduction}

The detection of weak signals in noise requires a careful analysis of the data with appropriate statistical tools. Given its simple use, one of the most popular
approaches is the spectral analysis of the time series by means of the periodogram. In astronomical applications
the use of this technique is not straightforward. In fact, from the statistical point of view this tool performs at its best only when the signals are sampled on a regular time grid, an uncommon situation in 
many astronomical observations. The analysis of a
periodogram in the case of irregular sampling is limited by the possibility to fix completely its statistical properties. This problem is not new \citep[e.g. see ][]{got75} and there have been
many attempts to solve it. The solution found in the Lomb-Scargle periodogram (LSP) approach \citep{lom76, sca82} comes at the price of theoretical difficulties which make its use not obvious.
In particular, the value of the LSP at a given frequency corresponds to what can be obtained by means of the least-squares fit of a single sine function with the same frequency \citep{sca82}.
This implies that the LSP is equivalent to a least-square fit of a temporal series using only a specific sine function at a time, i.e. it does not perform a simultaneous fit over the sine functions corresponding to all the inspected frequencies as it should be necessary.
However, only in the case of even samplings and in correspondence to the Fourier frequencies a set of sine functions constitutes an orthonormal basis and only in this case the fit of a single sine function
provides the same result as the simultaneous fit of the entire set of sine functions \citep[e.g. see ][p. 450]{ham73}.
This does not happen for uneven samplings. As a consequence the LSP it is not based on the correct least-squares model  (see appendix~\ref{sec:appC} for a formalized explanation).
Therefore, the real effectiveness and reliability of the implicit operation at the basis of the LSP is not founded on a correct ground.

Over the years many papers have been dedicated to understand and resolve the shortcomings of the LSP as, for example, the impossibility to describe it in terms of the spectral window, the development of specific computer codes, 
or the fact that, contrary to the even sampling case, the effect of the 
mean value of the signals is not limited to the zero frequency \citep[some recent works are][and reference therein]{ree07, zec09, vio10, vio13, van18}.  
However, an overlooked point is that ,although in the case of signals constituted only by noise the two periodograms provide almost identical results, the same may be not true if the searched signal is present. The consequence is that the two approaches provide different statistical significance to the presence of a signal  in a given time series. There are no compelling theoretical arguments that impose the use of the LSP instead of the classical periodogram (CP). Indeed, this last is
directly connected to the correlation function of the observed signal with the sinusoidal functions at the various frequencies of interest and hence
provides a direct estimate of the power spectrum avoiding the modeling of the data. This is even more true since, as it will be shown below, there are situations where the LSP is less effective than the CP in the detection of signals in noise.

In Sect.~\ref{sec:math} we discuss the limitations of the LSP and in Sect.~\ref{sec:new}  we describe the statistical characteristics of the entries of the CP. In 
Sect.~\ref{sec:cvlsp} the characteristics of the LSP and the CP are compared and tested in
Sect.~\ref{sec:test} with some numerical experiments, whereas in Sect.~\ref{sec:data} an experimental signal is outlined. Finally, Sect.~\ref{sec:conclusions} summarizes our conclusions.  For ease of reading three appendices have been added which contain most of the mathematics used in Sects.~\ref{sec:math}, ~\ref{sec:cvlsp}, and ~\ref{sec:test}.

\section{Mathematics of the problem} \label{sec:math}

When a continuous signal $x(t)$ is sampled on an irregular time grid $[t_0, t_1, \ldots, t_{M-1}]$,  a time series $[x_{t_0}, x_{t_1}, \ldots, x_{t_{M-1}}]$ is obtained. The CP of $\{x_{t_i}\}$ at a given frequency $\nu$ is defined as
\begin{equation} \label{eq:pf}
p_\nu = a_\nu^2 + b_\nu^2
\end{equation}  
where
\begin{align}
a_\nu & = \frac{1}{\sqrt{M}} \sum_{i=0}^{M-1} x_{t_i} \cos{2 \pi \nu t_i}, \label{eq:aki} \\
b_\nu & = \frac{1}{\sqrt{M}} \sum_{i=0}^{M-1} x_{t_i} \sin{2 \pi \nu t_i}. \label{eq:bki}
\end{align}
The function $p_\nu$ can be interpreted as the correlation between the time series and the sine and cosine modes with frequency $\nu$. It provides a statistical measure of the similarity between the time series
and a discrete sinusoidal signal of frequency $\nu$. Hence, a high peak in $p_{\nu}$  is indicative of the presence of a sinusoidal component of frequency $\nu$ in $\{ x_{t_i} \}$. For this reason,
the periodogram is typically used to test whether a time series contains the contribution of a deterministic signal $\{ s_{t_i} \}$ or it is constituted by a noise $\{ n_{t_i} \}$ only, i.e. to decide between the hypotheses $x_{t_i} = s_{t_i} + n_{t_i}$ and $x_{t_i} = n_{t_i}$.

In the case of a regular sampling time grid, i.e. $t_0 = 0, t_1 = 1, \ldots, t_{M-1}=M-1$, the question is
relatively simple if the periodogram is computed in correspondence to the so called Fourier frequencies, i.e. $\nu_k = k/(M-1)$, $k=0, 1, \ldots, M-1$. Indeed, under the hypothesis $x_i = n_i$, $i=0, 1, \dots, M-1$, with $n_i$ a zero-mean, 
Gaussian, stationary, white-noise process with standard deviation $\sigma_n$,
it can be verified that, independently of $k$, the normalized version $p_k/ \sigma_{n}^2 \to p_k$ is given by the sum of two  squared independent, zero-mean, Gaussian, random quantities with variance equal to $1/2$.
As a consequence, the corresponding probability density function (PDF) is the exponential distribution \citep{sca82},
\begin{equation}
g_{P_k}(p_k)= \exp{(-p_k)},
\end{equation}
with cumulative distribution function (CDF),
\begin{equation} \label{eq:Gp}
G_{P_k}(p_k)= 1-\exp{(-p_k)}.
\end{equation}
Hence, the probability of false alarm (PFA), that is the probability $\alpha$ with which at a generic frequency $p_k$ is expected to exceed by chance a level $\zf$, is
\begin{align}
\alpha &=1-G_{P_k}(\zf);  \label{eq:pfaG} \\
&= 1 - \left( 1 - {\rm e}^{-\zf} \right).  \label{eq:pfaG1}
\end{align}
However, if we consider all the frequencies of the periodogram, the probability $\alpha$ that one among them exceeds by chance a level $\zf$ is
\begin{align} 
\alpha &=1- G^{N^*}_{P_k}(\zf);  \label{eq:pfaGN} \\
&= 1 - \left( 1 - {\rm e}^{-\zf} \right)^{N^*},  \label{eq:pfaG1N}
\end{align}
with $N^*$ the number of statistically independent frequencies. Considering that whenever $k \ne k'$, with $k, k'=0, 1, \ldots M/2$, $p_k$ is independent of 
$p_{k'}$,  in the present case it is $N^*=M/2$. We call the probability $\alpha$ given by Eqs.~\eqref{eq:pfaGN} and \eqref{eq:pfaG1N} the specific (to the considered frequency)
probability of false alarm (SPFA). Using the SPFA it is possible to fix a detection threshold,
\begin{equation} \label{eq:threshold}
L_{{\rm Fa}} = - \ln\left[1 - (1 - \alpha)^{1/{N^*}} \right]
\end{equation}
which is called the ``level of false alarm''. 
Without loss of generality, here and henceforth $\sigma_n$ will be taken  to be unity \footnote{This can be obtained by standardize the time series to unit-variance, i.e. $x_i \to x_i/\sigma_n$}.

When the sampling is irregular, the situation becomes more complex given that $a_\nu$ and $b_\nu$ are zero-mean, Gaussian, random quantities but in this case correlated and with different variances. Hence, for a fixed frequency, the PDF of $p_\nu$ is not  given by $g_{P_{\nu}}(p_\nu)$. \citet{sca82} bypasses this problem introducing a modified version $\hat{p}_\nu$ of the CP,
\begin{equation}
\hat{p}_\nu  = \frac{1}{2} (\hat{a}_\nu^2 +\hat{b}_\nu^2),
\end{equation}
where
\begin{align}
\hat{a}_\nu & = \frac{\sum_{i=0}^{M-1} x_{t_i} \cos{[2 \pi \nu (t_i - \tau)]}}{\sqrt{\sum_{i=0}^{M-1} \cos^2{[2 \pi \nu (t_i - \tau)}}]}, \label{eq:ah} \\ 
\hat{b}_\nu & = \frac{\sum_{i=0}^{M-1} x_{t_i} \sin{[2 \pi \nu (t_i - \tau)]}}{\sqrt{\sum_{i=0}^{M-1} \sin^2{[2 \pi \nu (t_i - \tau)]}}}, \label{eq:bh}
\end{align}
with $\tau$ defined by
\begin{equation}
\tan{(4 \pi \nu \tau)} = \frac{\sum_{i=0}^{M-1} \sin{(4 \pi \nu t_i)}}{\sum_{i=0}^{M-1} \cos{(4 \pi \nu t_i)}}.
\end{equation}
The reason for such a modification is that, under the hypothesis $x_{t_i} = n_{t_i}$, $\hat{a}_\nu$ and $\hat{b}_\nu$ are zero-mean, unit variance, uncorrelated, Gaussian, random quantities. Therefore, the PDF of $\hat{p}_\nu$ is again of exponential type. 

A further problem which raises in the case of irregular sampling is that it is no longer possible to define a set of natural frequencies on which to compute the periodogram  such as the Fourier frequencies which permit the
reconstruction of the signal via the discrete inverse Fourier transform. Hence, there is no reason why the number $N$
of frequencies where to compute $\hat{p}_\nu$ must be equal to the number $M$ of the sampling time instants. Indeed, often $N \gg M$. Here, the point is that  the number $N^*$ of independent frequencies is not defined and the threshold $L_{{\rm Fa}}$ as given by Eq.~\eqref{eq:threshold}
is not computable. An analysis of this subject is available in \citet{hor86} and  \citet{van18} and references therein. However, as discussed in \citet{pre92}, the exact value of $N^*$ is not critical. Anyway, in the case of a semi-regular sampling 
and a regular frequency grid with step equal to  $1/ \Delta t$, with $\Delta t $ the smallest time step of the sampling, $N^* = M/2$ represents a reasonable choice \citep[e.g. see][]{vio10, vio13}. With random sampling patterns  $N^*=M$ may be an appropriate choice  \citep{pre92}. A more precise, although computationally expensive, estimation of $N^*$ can be obtained from the rank of the correlation matrix $\bf{\Sigma}$ of the periodogram entries (see appendix~\ref{sec:appB}). Recently, an alternative technique based on the extreme value theory has been proposed by \citet{bal08, bal14} \citep[see also][]{suv14, suv15}.

From the discussion above, it could be derived that with the LSP the computational issue of the PFA has been exactly solved, but in reality this is not the case.
Equations~\eqref{eq:ah} and \eqref{eq:bh} represent a local operation, i.e. an operation which is applied only on a frequency at a time and using specific normalization coefficients for that frequency, it does not use the data related to the remaining frequencies of interest and hence it operates differently at different frequencies. The consequence is that $\hat{p}_\nu$ represents a distorted version of $p_\nu$.
From both the theoretical and practical points of view this is not desirable and may give rise to unexpected consequences. In particular, as shown in Sect.~\ref{sec:new}, each entry of the CP 
has its own PDF $f_{P_{\nu}}(p_{\nu})$ and CDF $F_{P_{\nu}}(p_{\nu})$ which
are different for different frequencies. When $p_{\nu}$ is converted into $\hat{p}_{\nu}$, in general it results that $G_{\hat{P}_{\nu}}(\hat{p}_{\nu}) \neq F_{P_{\nu}}(p_{\nu})$. Hence, the same holds for the corresponding PFAs (see below). Given that the LSP is computed through a procedure
that distorts  the CP and, as seen above, it is not based on a correct
least-squares model, there is no reason why the outcoming PFAs should be preferred to those produced by the CP which is directly connected to the correlation function of the observed signal $\xb$ 
with the sinusoidal functions at the various frequencies of interest.

\section{Statistical characteristics of the entries of the classical periodogram} \label{sec:new}

The PDF $f_{P_{\nu}}(p_{\nu})$ of $p(\nu)$ is equivalent to the PDF
$f_Z(z)$ of the random variable $Z=X_1^2+X_2^2$ with $X_1$ and $X_2$ zero-mean, Gaussian, random variables  with standard deviation $\sigma_1$ and $\sigma_2$, respectively, and correlation coefficient $\rho$.  This PDF is given by
\citep[][see also~ appendix \ref{sec:appA} for a slightly different form best suited for implementation]{sim06}
\begin{equation} \label{eq:fz}
f_Z(z) = \frac{1}{2 \sigma_1 \sigma_2 \sqrt{1-\rho^2}} \exp{\left[ -\frac{1}{4} (\beta^+ - \gamma^+) z \right]}   I_0 \left( \frac{1}{4} \gamma^+ z \right),
\end{equation}
with $z \ge 0$,
\begin{align}
\gamma^+ &=\frac{\left[ (\sigma_1^2 + \sigma_2^2)^2 - 4 \sigma_1^2 \sigma_2^2 (1-\rho^2) \right]^{1/2}}{\sigma_1^2 \sigma_2^2 (1-\rho^2)}, \\ 
\beta^+ & = \gamma^+ + \frac{\sigma_1^2+\sigma_2^2}{\sigma_1^2 \sigma_2^2 (1-\rho^2)},
\end{align}
and $I_0(.)$ the modified Bessel function of the first kind of zero order. 

The corresponding central moments are given by  \footnote{Symbol ${\rm E}[.]$ denotes the expectation operator.}
\begin{equation}
{\rm E}\left[ \left(Z-\overline{Z} \right)^k \right] = \sum_{i=0}^{k} \frac{k!}{i! (k-i)!} {\rm E}[Z^i].
\end{equation}
where ${\rm E}[Z^k] $
\begin{multline}
{\rm E}[Z^k] = \frac{2^{2 k +1} k! }{ \sigma_1 \sigma_2 \sqrt{1-\rho^2} \left(\beta^+-\gamma^+\right)^{k+1}} \\
\times {}_{\phantom{1}2}F_1\left[\frac{k+1}{2}, \frac{k}{2}+1; 1; \left(\frac{\gamma^+}{\beta^+-\gamma^+}\right)^2 \right],
\end{multline}
with ${}_{\phantom{1}2}F_1(.,.;.;.)$ the Gauss hypergeometric function. Cases of interest are $k=1$ and $2$ for which
\begin{equation}
{}_{\phantom{1}2}F_1\left[\frac{k+1}{2}, \frac{k}{2}+1; 1; x \right] = \left\{
\begin{array}{ll}
\frac{1}{(1-x)^{3/2}} & \text{if } k=1,\\
\frac{2 + x}{2(1-x)^{5/2}} & \text{if } k=2.
\end{array} \right.
\end{equation}

The corresponding  cumulative distribution function $F_Z(z)$ is
\begin{equation} 
F_Z(z) = 1 +  \exp{\left[ -\frac{1}{4} (\beta^+ - \gamma^+) z \right]} I_0 \left( \frac{1}{4} \gamma^+ z \right) - 2 Q_1(A,B), \label{eq:F}
\end{equation}
where $Q_1(.,.)$ is the Marcum Q-function with
\begin{align}
A &=\frac{\sqrt{\sigma_1^2 + \sigma_2^2 - 2 \sigma_1 \sigma_2 \sqrt{1 - \rho^2}}}{2 \sigma_1 \sigma_2 \sqrt{1-\rho^2}} \sqrt{z}; \\
B &=\frac{\sqrt{\sigma_1^2 + \sigma_2^2 + 2 \sigma_1 \sigma_2 \sqrt{1 - \rho^2}}}{2 \sigma_1 \sigma_2 \sqrt{1-\rho^2}} \sqrt{z}.
\end{align}
At first sight these functions could appear rather convoluted, but they are commonly used in statistics and engineering \citep[e.g. see][]{hel68, van01, sha17} and indeed they may be found in all main software packages.

In the present case, it is $z=p_\nu$, $\sigma_1=\sigma_{a_\nu}$, $\sigma_2=\sigma_{b_\nu}$, $\rho=\rho_\nu$ with
\begin{align}
\sigma^2_{a_\nu} & = \frac{1}{M} \sum_{i=0}^{M-1} \cos^2{(2 \pi \nu t_i)}, \\
\sigma^2_{b_\nu} & = \frac{1}{M} \sum_{i=0}^{M-1} \sin^2{(2 \pi \nu t_i)},
\end{align}
and \citep{vio13}
\begin{align} 
\rho_{\nu} & = \frac{{\rm E}[a_\nu b_\nu]}{\sigma_{a_\nu} \sigma_{b_\nu}}; \nonumber \\
&= \frac{\sum_{i=0}^{M-1} \sin{(4 \pi \nu t_i)}}{2 \sqrt{\sum_{i=0}^{M-1} \cos^2{(2 \pi \nu t_i)}} \sqrt{\sum_{i=0}^{M-1} \sin^2{(2 \pi \nu t_i)}}} \label{eq:corr}.
\end{align}
The PDF $f_{P_{\nu}}(p_\nu)$  and the CDF $F_{P_{\nu}}(p_\nu)$ permit a spectral analysis of the time series based on the CP.

\section{Classical vs. the Lomb-Scargle periodogram} \label{sec:cvlsp}

There are various reasons why working with the CP is more advantageous than with the LSP. First,
with $f_{P_{\nu}}(p_\nu)$  and $F_{P_{\nu}}(p_\nu)$ the computation of the statistical significance of a peak in the periodogram does not require that  this last is modified frequency 
by frequency as it happens if Eqs.~\eqref{eq:ah}
and \eqref{eq:bh} are used. This makes the detection procedure more transparent and intuitive. 
Second, the CP does not require specific software for its implementation. Nonuniform fast Fourier transform algorithms are already widely available  \citep{bag99, pot01}. Moreover,
as shown in \citet{vio13}, if the time series are rebinned into a sufficiently high number of equispaced bins, the periodogram can be computed with arbitrary accuracy by means of the standard fast Fourier transform, with obvious computational gains.
Finally, the computation of the spectral window $w(\nu)$ does not present particular difficulties while, strictly speaking, it cannot be computed in the context of the LSP method. 
This is because, as seen above, the decorrelation of the coefficients $a_{\nu}$ and $b_{\nu}$ is a local operation, hence the mutual relationship among the various frequencies in $\hat{p}_{\nu}$ is altered.
We recall that the spectral window, defined as
\begin{equation}
w(\nu)= \left| \sum_{i=0}^{M-1}  {\rm e}^{- \iota 2 \pi \nu t_i} \right|^2,
\end{equation} 
with $\iota = \sqrt{-1}$, is key since all spectral leakage effects (aliasing, sidelobes, interference phenomena, ghosts, etc.) are manifested directly in it and can be easily evaluated quantitatively  \citep{dee75, sca82}.

With the CP, however, the detection threshold cannot be obtained in analytical
form. Therefore, in the case it is necessary to test the statistical significance of a peak at a prefixed frequency (e.g. when it is  known a priori that if a periodic signal of interest is present in a time series it contributes to the periodogram at a specific frequency)
the detection threshold must be numerically computed by solving the equation
\begin{equation} \label{eq:alpha}
1-F_{P_{\nu}}(L_{\rm Fa})=\alpha
\end{equation}
for $L_{\rm Fa}$. However, since $F_{P_{\nu}}(p_{\nu})$ is a monotone increasing function, its computation is not difficult.
An effective alternative is to fix a threshold value $\alpha$ for the PFA, to compute the quantities $\{ \alpha'_i \}$ for the peak of interest  $\{ p^*_{\nu_i} \}$,
\begin{equation} \label{eq:spfa}
\alpha'_i=1- F_{P_{\nu}}(p^*_{\nu_i}),
\end{equation}
and then to claim the peak as statistically significant if $\alpha'_i \leq \alpha$.

When the frequencies where the contribution of a signal takes place
are unknown, it is common practice to test the statistical significance of the highest peak $\zf_{\max}$ in the spectrogram.
However, its SPFA cannot be computed on the basis of the method used to obtain Eq.~\eqref{eq:pfaG1N} since each entry of $p_{\nu}$ has its own PDF.
This is the main reason at the basis of the LSP where all entries are forced to follow the same exponential distribution. Only in this way the SPFA can be estimated by means of
Eq.~\eqref{eq:pfaG1N} or the technique by \citet{bal08, bal14}.  A possible approach to avoid this issue is based on the fact
that in most cases of interest the coefficients $a_{\nu}$ and $b_{\nu}$ are almost uncorrelated for the vast majority of the frequencies \citep{sca82, vio10,vio13}. 
Hence, their PDFs are expected to be similar and then the SPFA corresponding to $\zf_{\max}$ can be computed on the basis of
\begin{equation} \label{eq:Fz}
\alpha=1-\bar{F}_{\Zf_{\max}}^{N^*}(\zf_{\max}),
\end{equation}
where the CDF $\bar{F}_{P_{\nu}}(p_{\nu})$, with PDF $\bar{f}_{P_{\nu}}(p_{\nu})$, is given by Eq.~\eqref{eq:F} with $\sigma_{a_\nu}$, $\sigma_{b_\nu}$ and $\rho_{\nu}$ substituted, respectively, by $\bar{\sigma}_{a}$, $\bar{\sigma}_{b}$ and $\bar{\rho}$.
The first two quantities correspond to the mean values of $\sigma_{a_\nu}$ and $\sigma_{b_\nu}$ evaluated on all the frequencies of the periodogram whereas the last quantity 
corresponds to the mean value of the absolute values of  $\rho_{\nu}$.
Alternatively and more simply, it is possible to exploit the fact that, as shown by \citet{vit96a, vit96b} on the basis of a theoretical study using the spectral window, the CP is typically
almost identical to the LSP. As a consequence, the entries of $p_{\nu}$ are approximately distributed according to the classical exponential PDF. In other words, the SPFA of a peak in $p_{\nu}$ can be estimated on the basis of Eq.~\eqref{eq:pfaG1N} exactly as it is done for the LSP. This approach is suggested, although with some caution, also by \citet[][Sect.~4.4]{bal14}.

Here one could ask why do not use directly the LSP. 
The point is that, as it is shown in appendix~\ref{sec:appA}, when $\rho$ is sufficiently large,
$f_Z(z)$ given by Eq.~\eqref{eq:fz} can be well approximated by
\begin{equation}
f_Z(z) \approx \frac{1}{\sqrt{2 \pi z}} {\rm e}^{-z/2}.
\end{equation}
This function decreases with $z$  slower  than the classical exponential distribution.
With the LSP approach the entries of the periodogram $p_{\nu}$ are modified and forced to follow the exponential distribution which therefore has a faster right tail decay than $f_{P_{\nu}}(p_{\nu})$.  
The consequence is that the peaks which in $p_{\nu}$ are related to frequencies characterized by $|\rho_{\nu}| \gg 0$ in $\hat{p}_{\nu}$ tend to be more lowered than the peaks characterized by $\rho_{\nu} \approx 0$, i.e. in the LSP the relative heights of the peaks are modified. 
Such effect becomes important for high values of $p_{\nu}$. The consequences are potentially harmful since a peak due to a signal well above the level of the noise in the CP
can be below the noise level in the LSP (see below for an example).
For this reason, if $|\rho_{\nu}| \approx 0$ for the frequency of the peak of interest, its SPFA  can be computed by means of the CP and one of the above mentioned methods. In the improbable case that $|\rho_{\nu}| \gg 0$ some further validation is necessary (e.g., by means of numerical simulations).

\section{Numerical examples} \label{sec:test}

\begin{figure}
\vskip -3cm
        \resizebox{\hsize}{!}{\includegraphics{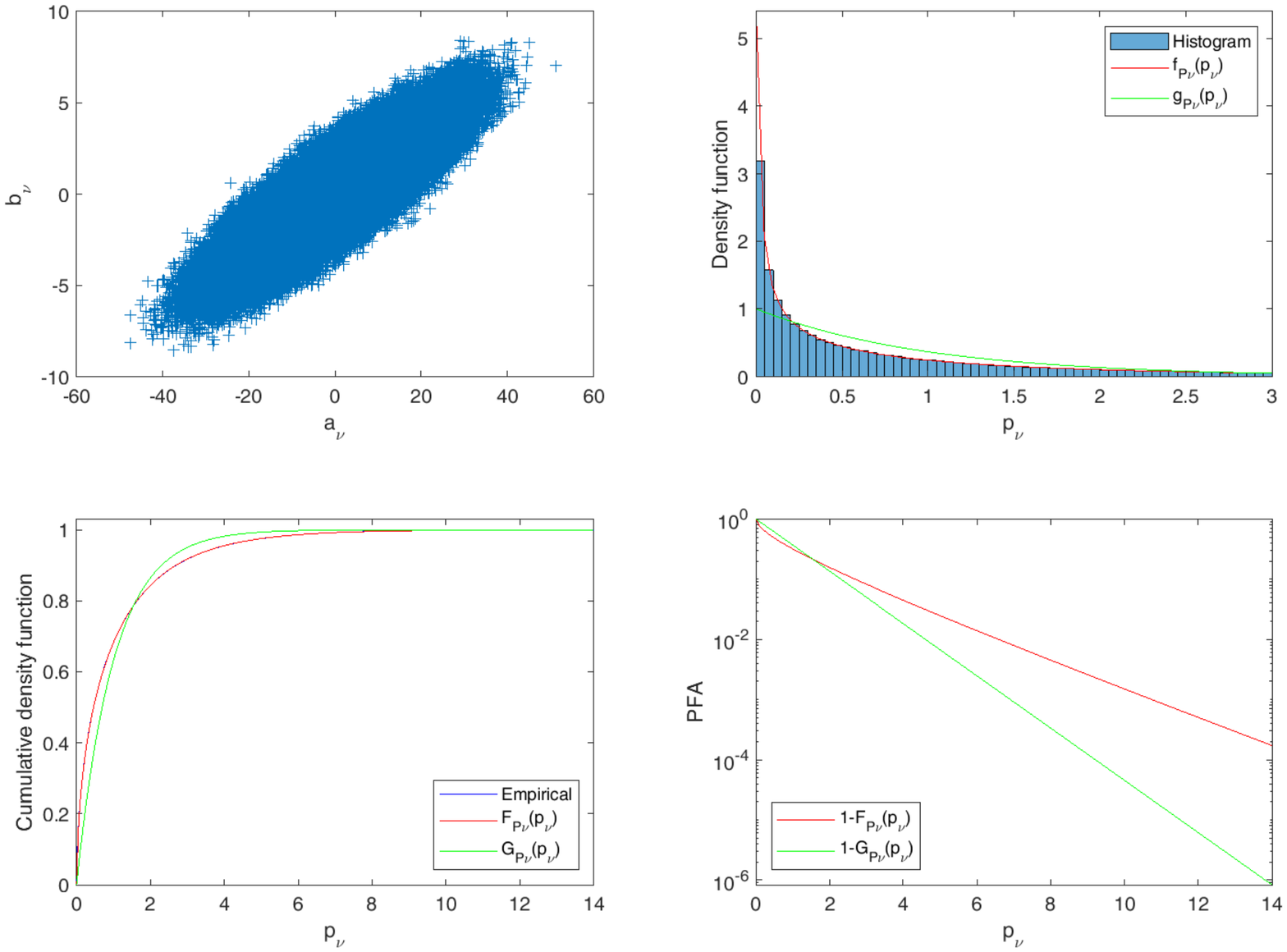}}
\vskip -1cm
        \caption{Top-left panel: coefficients $a_\nu$ vs. $b_\nu$, when $\nu=0.05$,  for the numerical experiment where $10^6$ zero-mean, unit-variance, Gaussian, white-noise processes are simulated on a time sampling grid containing
 periodic gaps (see the text). Note the strong correlation between these two quantities due to the high values of the correlation coefficient $\rho_{\nu}=0.85$. Top-right and bottom-left panel: PDF $f_{P_{\nu}}(p_\nu)$ and the CDF $F_{P_{\nu}}(p_\nu)$ of the values of the periodogram $p_{\nu} = a_{\nu}^2 + b_{\nu}^2$. For reference, the exponential PDF $g_{P_{\nu}}(p_\nu)$ and CDF $G_{P_{\nu}}(p_\nu)$ are plotted as well as the histogram and the empirical 
CDF. The almost perfect overlapping of  $f_{P_{\nu}}(p_\nu)$ and $F_{P_{\nu}}(p_\nu)$  with the corresponding empirical functions is clear. Indeed, they are indistinguishable in the figures. The same is not true for  $g_{\hat{P}_{\nu}}(\hat{p}_\nu)$ and  
$G_{P_{\nu}}(p_\nu)$. Bottom-right panel: the probability of false alarm $1- F_{P_{\nu}}(p_\nu)$ vs. the probability of false alarm $1-G_{P_{\nu}}(p_\nu)$.}.
        \label{fig:fig_result005}
        \resizebox{\hsize}{!}{\includegraphics{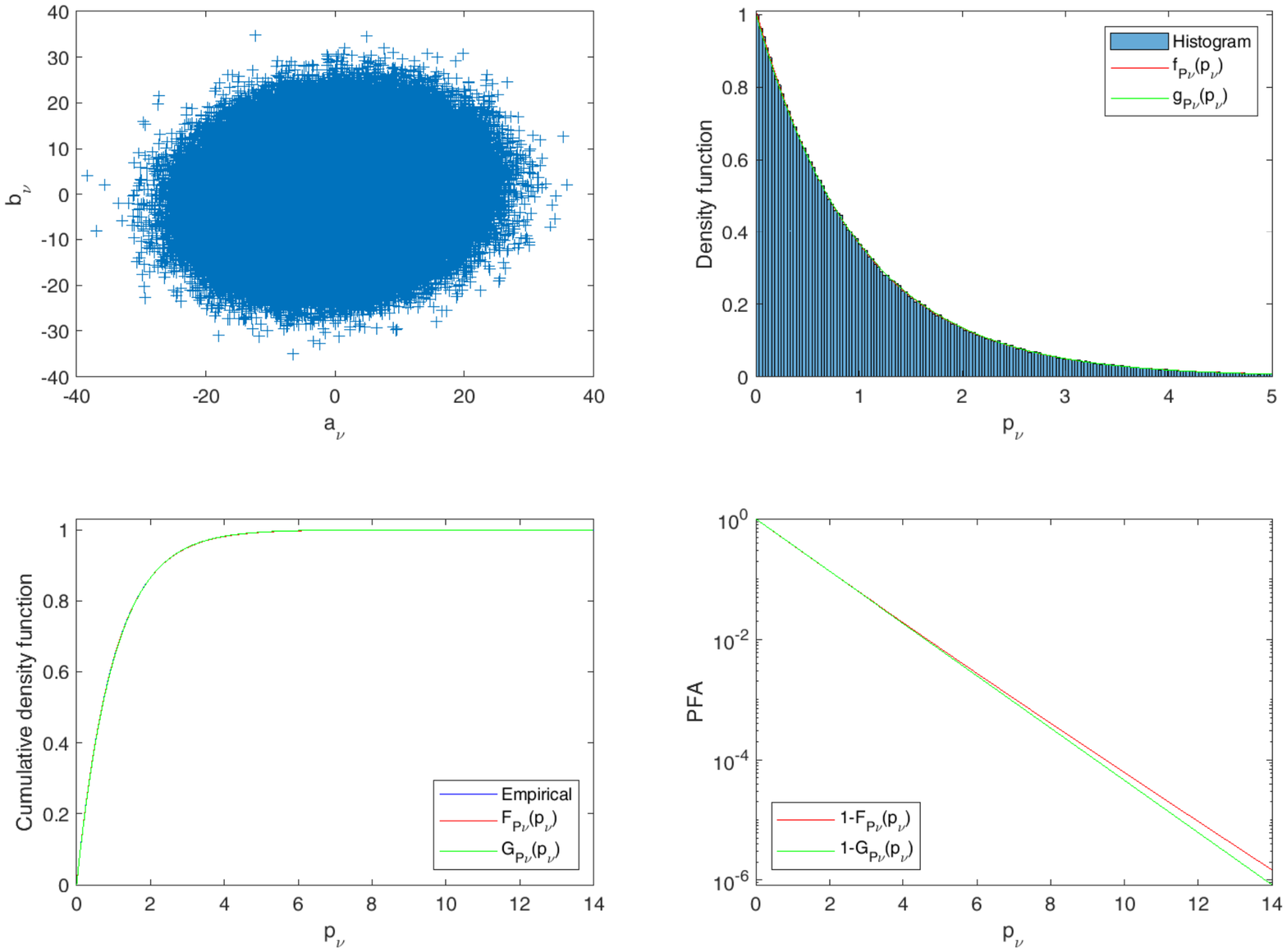}}
\vskip -1cm
        \caption{The same as in Fig.~\ref{fig:fig_result005} but for $\nu=1.20$. Here, $g_{P_{\nu}}(p_\nu)$ and $G_{P_{\nu}}(p_\nu)$ are very similar to $f_{P_{\nu}}(p_\nu)$ and $F_{P_{\nu}}(p_\nu)$, respectively. This is the consequence
of the small value of the correlation coefficient $\rho_{\nu} \approx 0$.}
        \label{fig:fig_result120}
\end{figure}

Figures~\ref{fig:fig_result005} and \ref{fig:fig_result120}  show the relationship between the coefficients $a_{\nu}$ and $b_{\nu}$ at two specific frequencies, as well the corresponding
$f_{P_{\nu}}(p_\nu)$  and $F_{P_{\nu}}(p_\nu)$, resulting from a numerical experiment where $10^6$ realizations of a discrete zero-mean, unit-variance, Gaussian, white-noise process is 
simulated with sampling time instants given by $t_i = (m \times 5 +  j)/205$, 
$i=j +m/2$ with $j=1,2,\ldots, 5$, and $m=0, 2, \ldots, 40$. The resulting time series contain $105$ data with sampling 
times in the range $[0, 1]$ where every sequence of five times containing an observation is followed by a 
sequence of five times with no data. In this way, a time series with periodic gaps is simulated. The sampling time step of the data is  $\Delta t = 4.902 \times 10^{-3}$ with corresponding Nyquist frequency $\nu_{\rm Ny} = 102$.
In practical applications time series similar to these ones are not very common. However,  the perfect regularity of both the sampling and the gap distribution exacerbates certain problems related to the irregular sampling and hence make easier 
their analysis.

Two frequencies are considered, $\nu=0.05$ and $\nu=1.20$. Since the frequency resolution of a periodogram is provided by the inverse of the time span $\Delta T = t_{\max}-t_{\min}$ of the input data (equal to one in the present case),
the first frequency corresponds to a sinusoid which is not sampled on a complete cycle. The reason for such a choice is that, as it is visible in top-left panel of Fig.~\ref{fig:fig_result005}, the corresponding coefficients $a_{\nu}$ and $b_{\nu}$ are 
strongly correlated ($\rho_{\nu}=0.85$). For reasons explained below, this is a situation difficult to reproduce with frequencies greater than $1/ \Delta T$. 
Indeed, $\rho_{\nu}$ results much less significant for the frequencies $\nu \ge 1$. An example is given in the top-left panel of Fig.~\ref{fig:fig_result120} for the frequency $\nu=1.20$ for which $\rho_{\nu}=0.17$.

The top-right and the bottom-left panels of Fig.~\ref{fig:fig_result005} show that $f_{P_{\nu}}(p_\nu)$  and $F_{P_{\nu}}(p_\nu)$ at $\nu=0.05$ are different from the corresponding exponential counterpart $g_{P_{\nu}}(p_\nu)$ and $G_{P_{\nu}}(p_\nu)$,
but only the first two are in good agreement with the respective empirical estimators of the PDF and the CDF.  The bottom-right panel in the same figure shows that the PFAs provided by the two CDFs  are very different. As the corresponding panels in Fig.~\ref{fig:fig_result120} show, these differences almost disappear for the frequency $\nu=1.20$. 

\begin{figure}
\vskip -3cm
        \resizebox{\hsize}{!}{\includegraphics{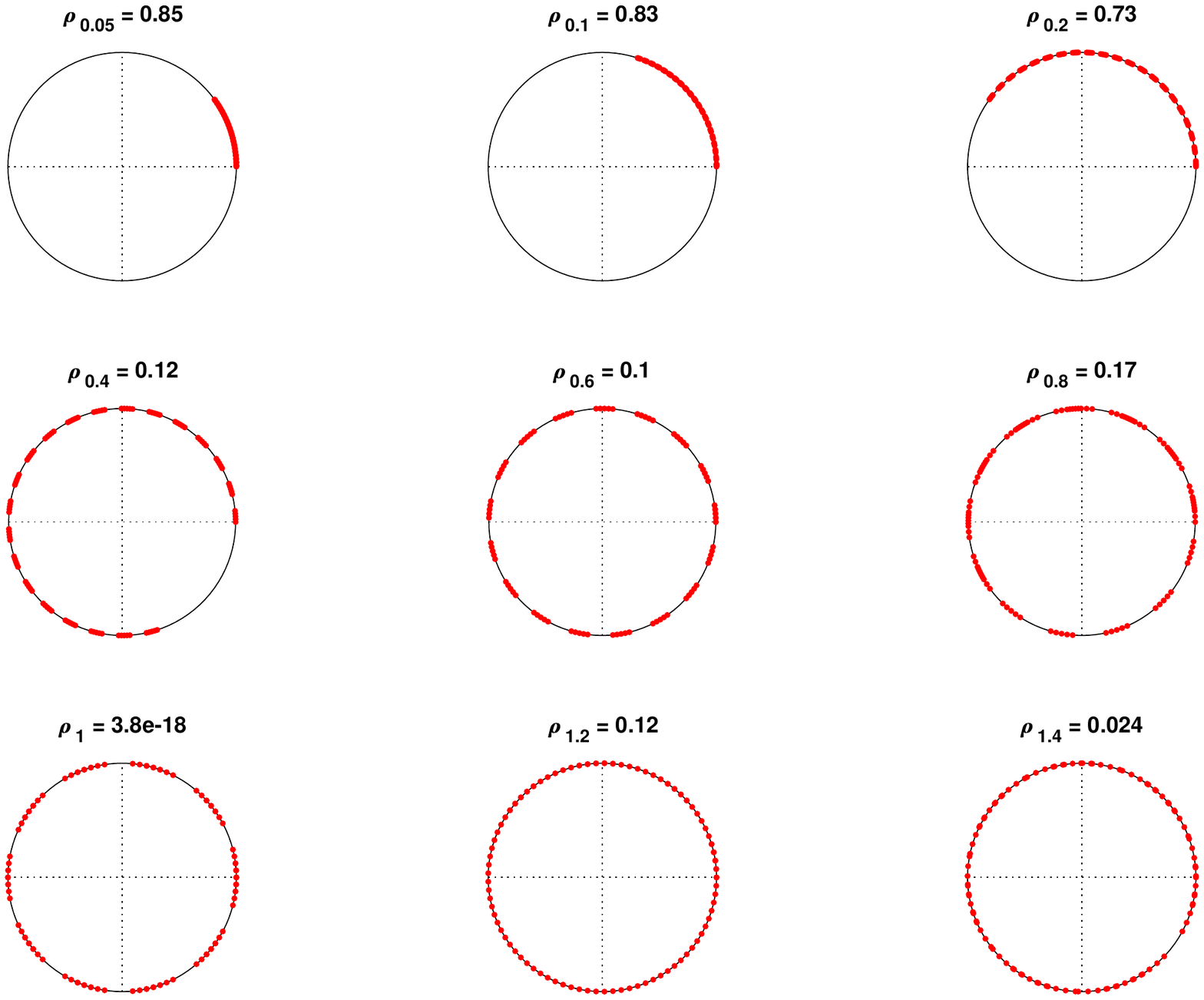}}
\vskip -1cm
        \caption{Distribution of the angles $\theta_i = 4 \pi \nu t_i$ on the unit circle corresponding to a set of different frequencies $\nu$ for the numerical experiments described in the text. The circles corresponding to the frequencies $\nu=0.05$ and
$1.20$ are to be related to the results shown in Figs.~\ref{fig:fig_result005} and \ref{fig:fig_result120}.}
        \label{fig:fig_circles}
        \resizebox{\hsize}{!}{\includegraphics{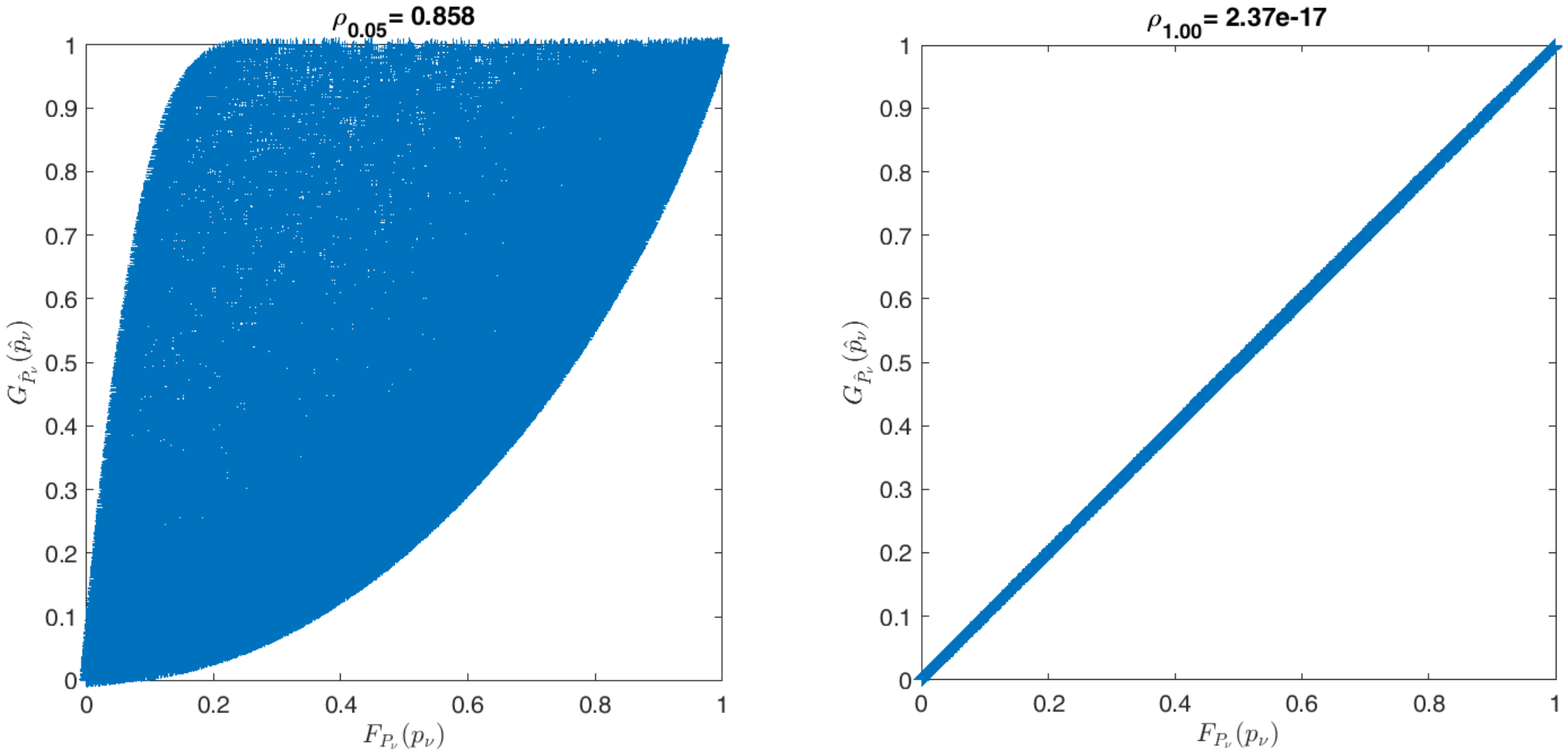}}
\vskip -1cm
        \caption{ CDF $F_{P_{\nu}}(p_{\nu})$ vs. CDF $G_{\hat{P}_{\nu}}(\hat{p}_{\nu})$ for the $10^6$ time series used in the Figs.~\ref{fig:fig_result005} and \ref{fig:fig_result120}. Here, the frequencies $\nu = 0.05$ and $\nu = 1.00$ have been used which have  $\rho_{\nu}$ equal to $0.85$ and approximately $0$, respectively. It is evident that, especially for great values of the $|\rho_{\nu}|$, the decorrelation of the coefficients $a_{\nu}$ and $b_{\nu}$ operated by the 
Lomb-Scargle method alters the statistical characteristics of the periodogram.}
        \label{fig:fig_pfa}
\end{figure}

These results are easy to understand by considering the distribution of the angles $\theta_i = 4 \pi \nu t_i$ of a unit circle.
Indeed, from Eq.~\eqref{eq:corr} it results that the correlation $\rho_\nu$ between the coefficients $a_\nu$ and $b_\nu$ at a given frequency $\nu$ is proportional to $\sum_{i=0}^{M-1} \sin{(4 \pi \nu t_i)}$. Since the sine function is an odd function, one may expect that $\rho_\nu \approx 0$ if
the angles $\{ \theta_i \}$ of a unit circle are uniformly and/or symmetrically distributed. This is a very probable situation for the vast majority of the frequencies. Indeed, even if for a given frequency it happens that the angles $\{ \theta_i \}$
cluster somewhere on the unit circle, it is quite improbable that such a cluster could survive for the other frequencies \citep{vio13}.
 The different distributions on the unit circle of the angles $\{ \alpha_i \}$ corresponding to the two frequencies $\nu=0.05$ and $\nu=1.20$ are visible 
in Fig.~\ref{fig:fig_circles}. It is clear that for the frequency $\nu=0.05$ the distribution is asymmetric contrary to that corresponding to the frequency $\nu=1.20$.
Similar experiments with other kinds of sampling time grids provide identical results. 

\begin{figure}
\vskip -3cm
        \resizebox{\hsize}{!}{\includegraphics{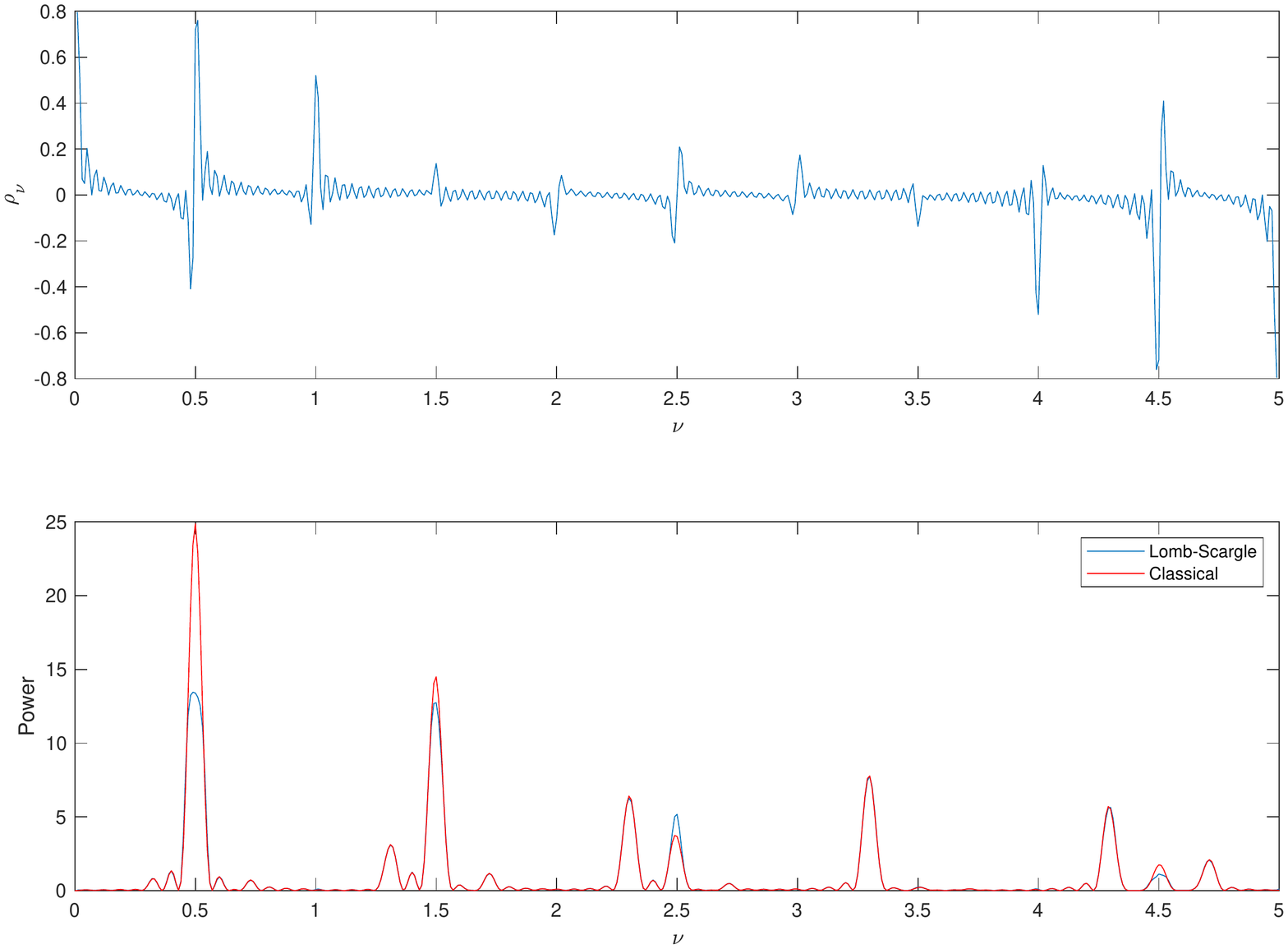}}
\vskip -1cm
        \caption{Top panel: Correlation coefficients $\rho_{\nu}$ between the coefficient $a_{\nu}$ and $b_{\nu}$  for the periodogram of a signal $x(t) = \cos{(2 \pi \nu_1 t)} + \cos{(2 \pi \nu_2 t)}$
 where the sampling time instants are regularly distributed with a constant time interval $\Delta t$ and to $\Nc_o$ successive observations follow $\Nc_m$ missing points
with the group of $\Nc_o + \Nc_m$ points repeated $\Mc$ times. Here, $\nu_1= 0.5$, $\nu_2=3.3$, $\Nc_o=3$, $\Nc_m=7$, $\Mc=15$ and $\Delta t = 0.1$.
Contrary to the frequency $\nu_2$, the frequency $\nu_1$ corresponds to a frequency of the CP with $\rho_{\nu} \gg 0$. Bottom panel: corresponding Lomb-Scargle  and CP.}
        \label{fig:fig_vityazev1}
        \resizebox{\hsize}{!}{\includegraphics{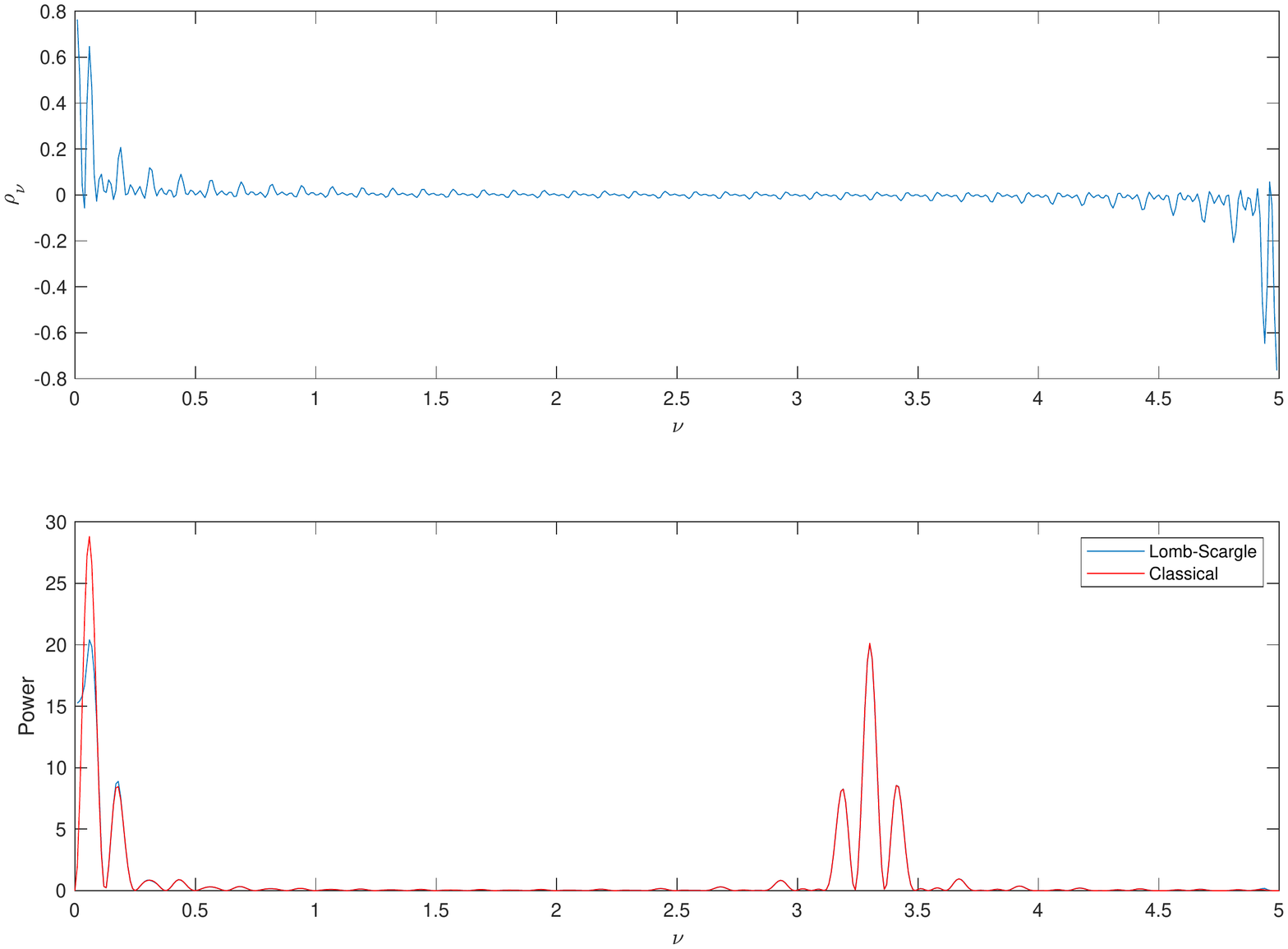}}
\vskip -1cm
        \caption{The same as in Fig.~\ref{fig:fig_vityazev1} but with $\nu_1=0.0625$, $\nu_2=3.3$, $\Nc_o=40$, $\Nc_m=40$, $\Mc=2$ and $\Delta t = 0.1$. This case correspond to a signal
with a sampling characterized by a long gap. Also here, contrary to the frequency $\nu_2$, the frequency $\nu_1$ corresponds to a frequency of the CP with $\rho_{\nu} \gg 0$.}
        \label{fig:fig_vityazev2}
\end{figure}

The two effects that a high value of $\rho_{\nu}$ can have on $\hat{p}_{\nu}$, mentioned at the end of Sects.~\ref{sec:math} and \ref{sec:cvlsp}, are visible in Figs.~\ref{fig:fig_pfa}-\ref{fig:fig_vityazev2}. In particular, 
Fig.~\ref{fig:fig_pfa} shows $G_{\hat{P}_{\nu}}(\hat{p}_{\nu})$ vs. $F_{P_{\nu}}(p_\nu)$
for two frequencies, $\nu=0.05$ and $\nu=1.00$, characterized by a correlation coefficient $\rho_{\nu}$ equal to $0.85$ and approximately $0$, respectively. This experiment is based on the $10^6$ time series
used above. It is evident that when $|\rho_{\nu}|$ is small the two CDF's are, as expected, identical. The situation drastically changes for greater vales of $|\rho_{\nu}|$. The result of this exercise is telling us that, as stated above, the PFA 
and hence the SPFA obtained from the LSP and the CP can be very different for high values of $|\rho_{\nu}|$.

The second effect is visible in the top panels of Figs.~\ref{fig:fig_vityazev1} and \ref{fig:fig_vityazev2} which show the CP and the LSP  (frequency range $[0, 5]$) for the signal used in the numerical experiments
by \citet{vit96a} where the signal
\begin{equation} \label{eq:signal}
x(t) = \cos{(2 \pi \nu_1 t)} + \cos{(2 \pi \nu_2 t)}
\end{equation}
is sampled with periodic gaps. In particular, the sampling time instants are regularly distributed with a constant time interval $\Delta t$ and to $\Nc_o$ successive observations follow $\Nc_m$ missing points
with the group of $\Nc_o + \Nc_m$ points repeated $\Mc$ times. The top panel in Fig.~\ref{fig:fig_vityazev1} shows the correlation coefficient $\rho_{\nu}$ for the simulation with
$\nu_1= 0.5$, $\nu_2=3.3$, $\Nc_o=3$, $\Nc_m=7$, $\Mc=15$ and $\Delta t = 0.1$ where
it is visible that, contrary to the frequency $\nu_2$, the frequency $\nu_1$ is characterized by a $\rho_{\nu} \gg 0$. The bottom panel in the same figure shows the corresponding LSP and CP. As expected,
the amplitude of the peak at the frequency $\nu_1$ in the LSP is smaller than the amplitude of the corresponding peak in the CP. The same does not hold for the peak at the frequency $\nu_2$.
In the experiment of Fig.~\ref{fig:fig_vityazev2},  which simulates observations characterized by a long gap, the signal \eqref{eq:signal}, with $\nu_1=0.0625$ and $\nu_2=3.3$, is sampled  on a time grid with $\Nc_o=40$, $\Nc_m=40$, $\Mc=2$ and 
$\Delta t = 0.1$. Also here it is $\rho_{\nu} \gg 0$
for the frequency $\nu_1$ whereas  $\rho_{\nu} \approx 0$ for the frequency $\nu_2$. Again, it is evident that in the LSP the peak at the frequency $\nu_1$ is lower than the corresponding peak in the CP. 

\begin{figure}
\vskip -3cm
        \resizebox{\hsize}{!}{\includegraphics{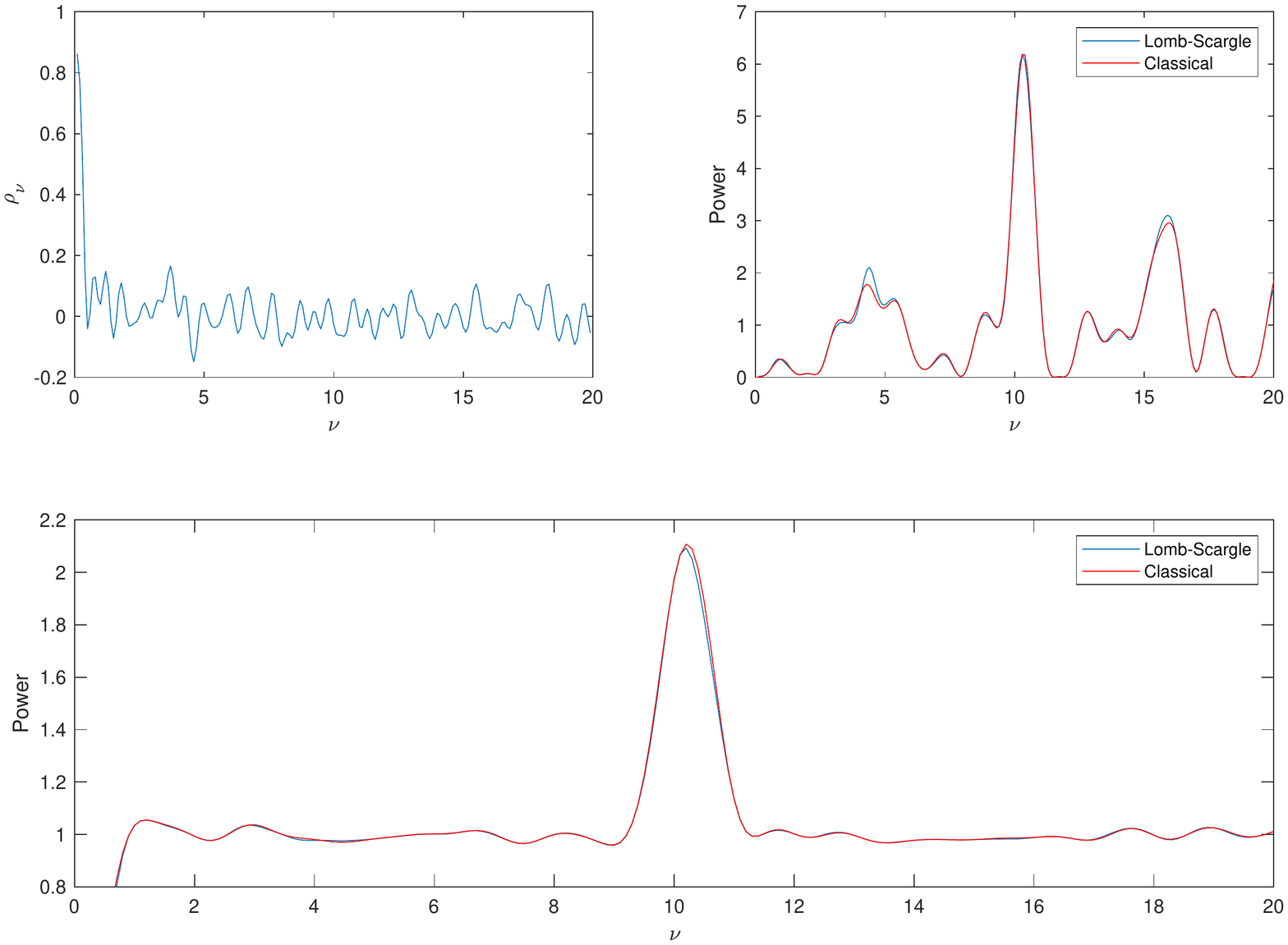}}
\vskip -1cm
        \caption{Top-left panel: Correlation coefficients $\rho_{\nu}$ between the coefficient $a_{\nu}$ and $b_{\nu}$  for the periodogram of a zero-mean, unit-variance, Gaussian, white-noise process
added with a sinusoidal signal of frequency $\nu=10.2$ and amplitude set to $0.15$. The sampling time grid of the signal is constituted by $200$ instants uniformly and randomly distributed in the interval $[0, 1]$. 
The  LSP and the CP are computed on $200$ frequencies uniformly distributed in the range  $[0, 20]$. Bottom-panel: mean of the Lomb-Scargle  and CP obtained from $5000$ times series with the same characteristics of the time series used in the top-right panel. To notice also in this case that, since $\rho_{\nu} \approx 0 $ for all the frequencies, the Lomb-Scargle and the CP are very similar.}
        \label{fig:fig_sin_rand}
        \resizebox{\hsize}{!}{\includegraphics{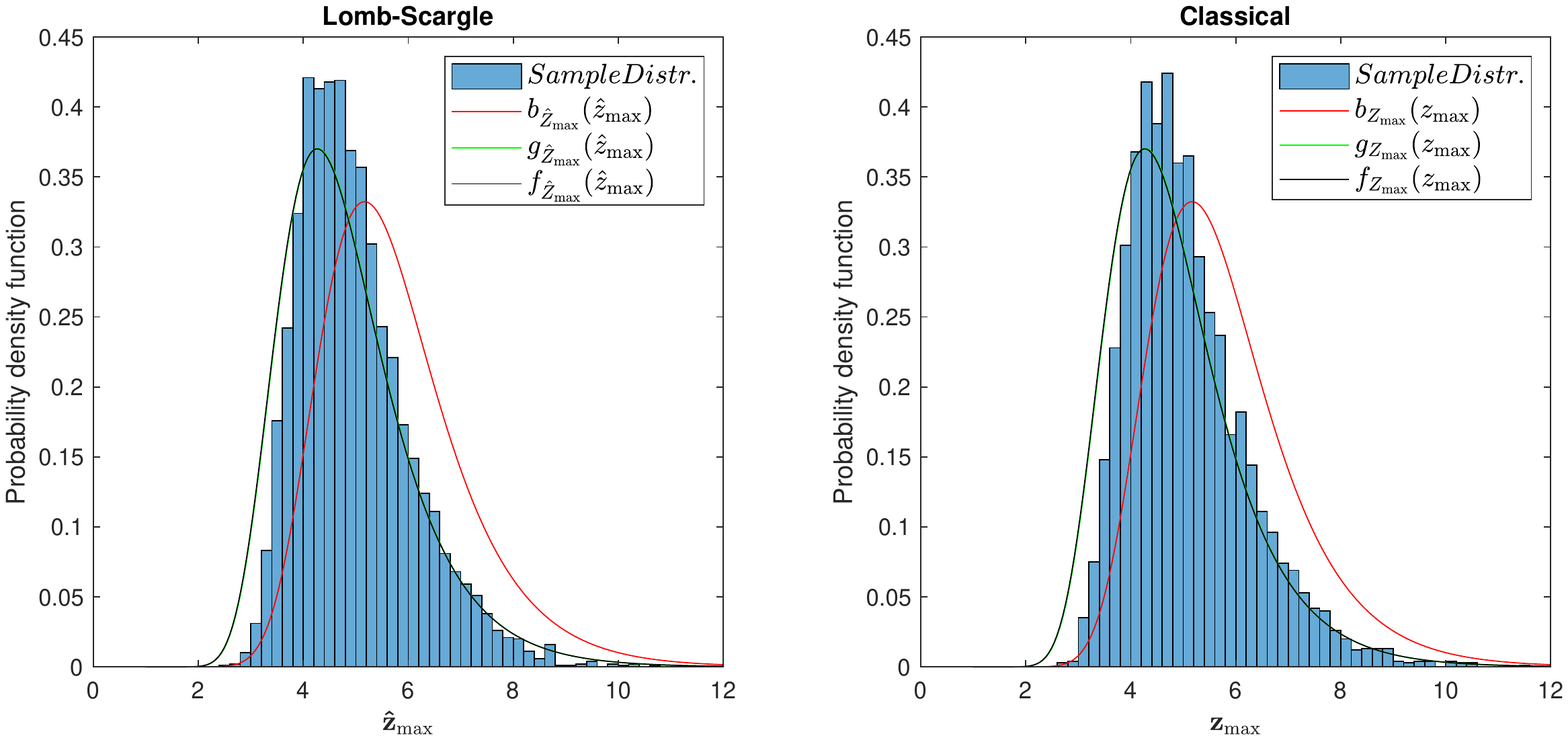}}
\vskip -1cm
        \caption{Histogram of the greatest peak for both the LSP and the CP obtained from $5000$ numerical simulations of the experiment in Fig.~\ref{fig:fig_vityazev1}
when the time series contains only a zero-mean, unit-variance white noise. The PDF $\bbf_{\hat{\Zf}_{\max}}(\hat{\zf}_{\max})$ derived from the extreme value theory by \citet{bal08}, the PDF 
$\gf_{\hat{\Zf}_{\max}}(\hat{\zf}_{\max})$ based on the exponential distribution given by Eq.~\eqref{eq:gpdf} and the proposed PDF $\ff_{\Zf_{\max}}(\zf_{\max})$ given by Eq.~\eqref{eq:gbpdf} are also plotted.}
        \label{fig:fig_SPFA1}
\end{figure}

\begin{figure}
\vskip -3cm
        \resizebox{\hsize}{!}{\includegraphics{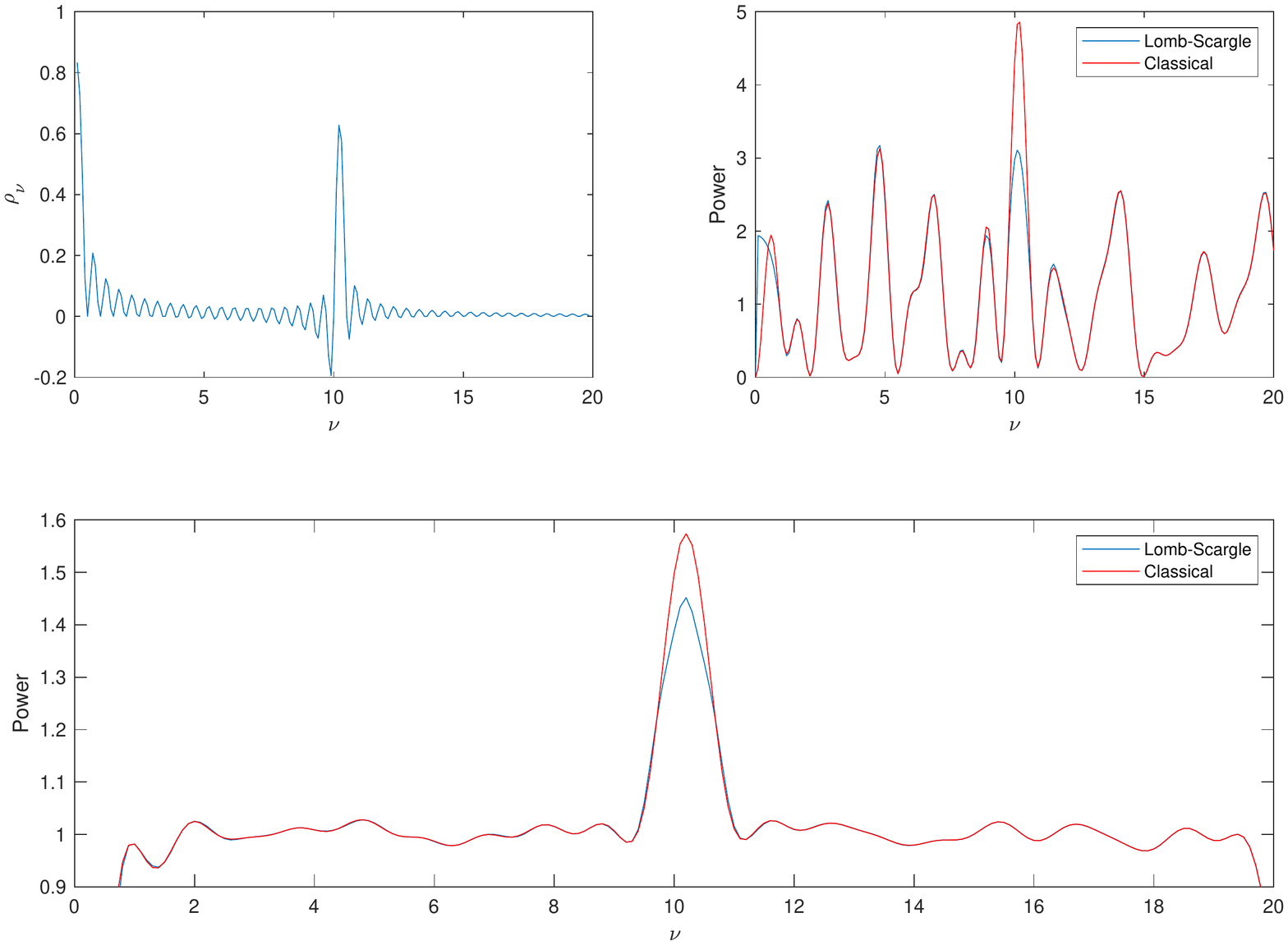}}
\vskip -1cm
        \caption{Top-left panel: Correlation coefficients $\rho_{\nu}$ between the coefficient $a_{\nu}$ and $b_{\nu}$  for the periodogram of a zero-mean, unit-variance, Gaussian, white-noise process
added with a sinusoidal signal of frequency $\nu=10.2$ and amplitude set to $0.15$, sampling time grid identical to that used in
 Figs.~\ref{fig:fig_result005} and \ref{fig:fig_result120} and computed on $200$ frequencies uniformly distributed in the range  $[0, 20]$. The frequency of the sinusoid corresponds to the frequency of the CP with the greatest correlation
coefficients $\rho_{\nu}$ (see text). Top-right panel: corresponding Lomb-Scargle  and CP. In the Lomb-Scargle periodogram it is visible the lowering of the highest peak with respect to the other peaks which makes it indistinguishable 
from the noise. Bottom-panel: mean of the Lomb-Scargle  and CP obtained from $5000$ times series with the same characteristics of the time series used in the top-right panel.}
        \label{fig:fig_sin}
        \resizebox{\hsize}{!}{\includegraphics{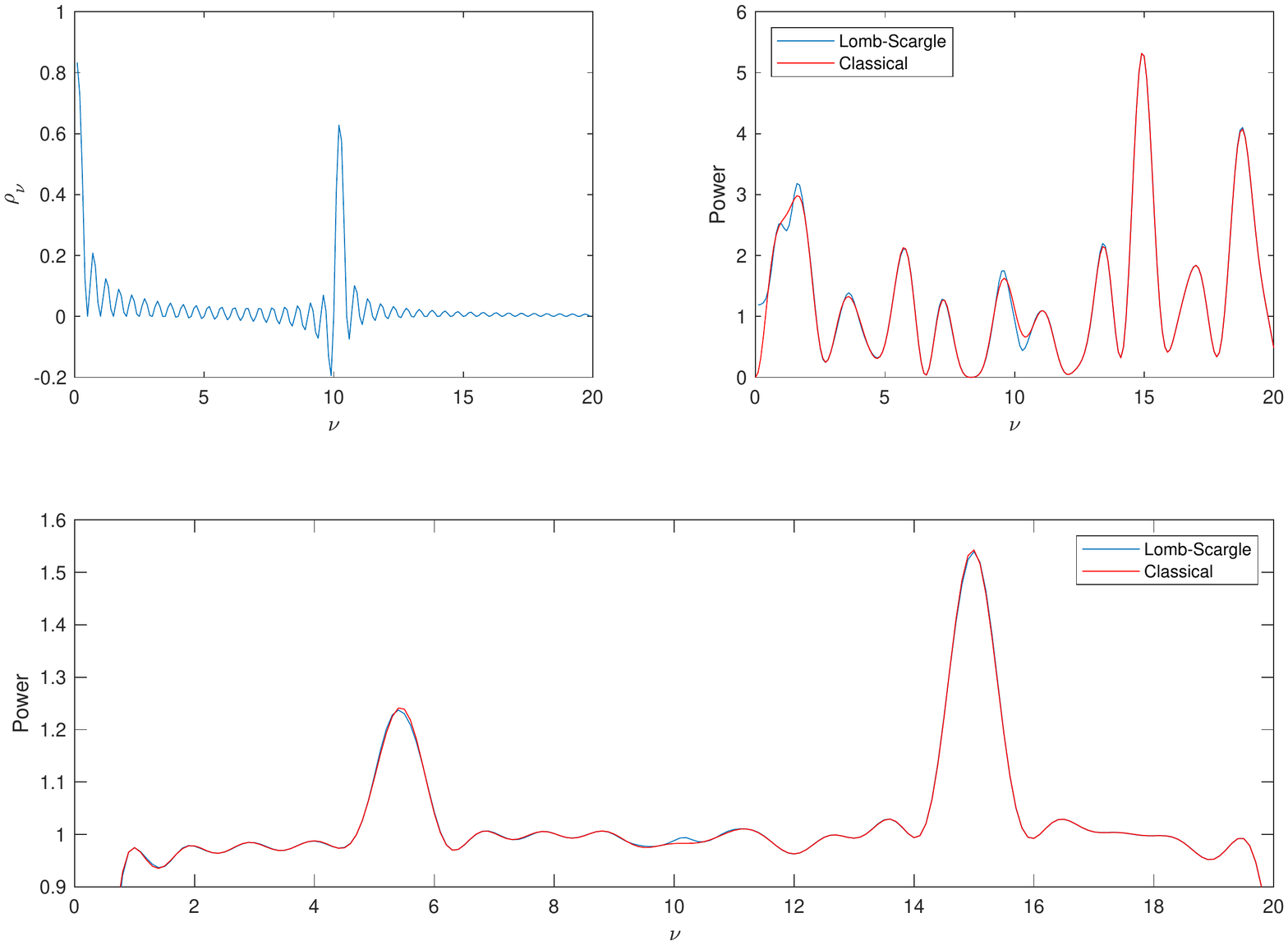}}
\vskip -1cm
        \caption{The same as in Fig.~\ref{fig:fig_sin} but with a sinusoidal signal of frequency $\nu=15.0$ and amplitude again set to $0.15$. This frequency is characterized by a correlation coefficient 
$\rho_{\nu} \approx 0$. Contrary to the previous figure, here the Lomb-Scargle  and CP are almost indistinguishable.}
        \label{fig:fig_sin_2}
\end{figure}

The lowering of the peaks in the LSP can have consequences in the case of noisy signals. This is visible in the top-right panel of Fig.~\ref{fig:fig_sin} which shows the CP and the LSP  (frequency range $[0, 20]$) of 
a simulated time series with the same characteristics of the signal used in Fig.~\ref{fig:fig_pfa} added to a sinusoidal signal. The amplitude of the sinusoid is $0.15$ whereas its frequency is $\nu=10.2$. This last corresponds to the frequency in the classical peridogram
with the greatest $\rho_{\nu}$ (see the top-left panel of the same figure). It is clear that while $p_{10.2}$ corresponds to a predominant peak at the frequency of the sinusoid, the same is not true for $\hat{p}_{10.2}$.
In other words, with the LSP it is not even possible to guess the presence of a signal in the time series. Again, this is the consequence of the above mentioned fact that when $|\rho_{\nu}| \gg 0$ 
the PDF $g_{\hat{P}_{\nu}}(\hat{p}_{\nu})$ is more short tailed 
than $f_{P_{\nu}}(p_\nu)$. That this is not a conclusion due to a single simulation is supported by the bottom-panel of Fig.~\ref{fig:fig_sin} which shows the mean of the CPs and of the LSPs obtained from $5000$ simulated time series with the same characteristics of the time series used for the top panels. As a countercheck, the same effect is not visible in Fig.~\ref{fig:fig_sin_2} where the same experiment is repeated with an amplitude 
of the sinusoid set again to $0.15$ but with a frequency $\nu=15.0$ for which $\rho_{\nu} \approx 0$. 

The same conclusion is supported by Fig.~\ref{fig:fig_sin_rand} which shows
what happens with the same signal as in the previous experiment when $200$ sampling time instants are randomly and uniformly distributed in the interval $[0, 1]$. 
These kinds of sampling is by far less-pathological than 
the one considered above. From these figures it is evident that $|\rho_{\nu}| \approx 0$ for all the considered frequencies. Hence, the LSP and the CP are, as expected, very similar. 
By the way, this is the reason why in practical applications the limitations of the LSP are not apparent.

\begin{figure}
\vskip -3cm
        \resizebox{\hsize}{!}{\includegraphics{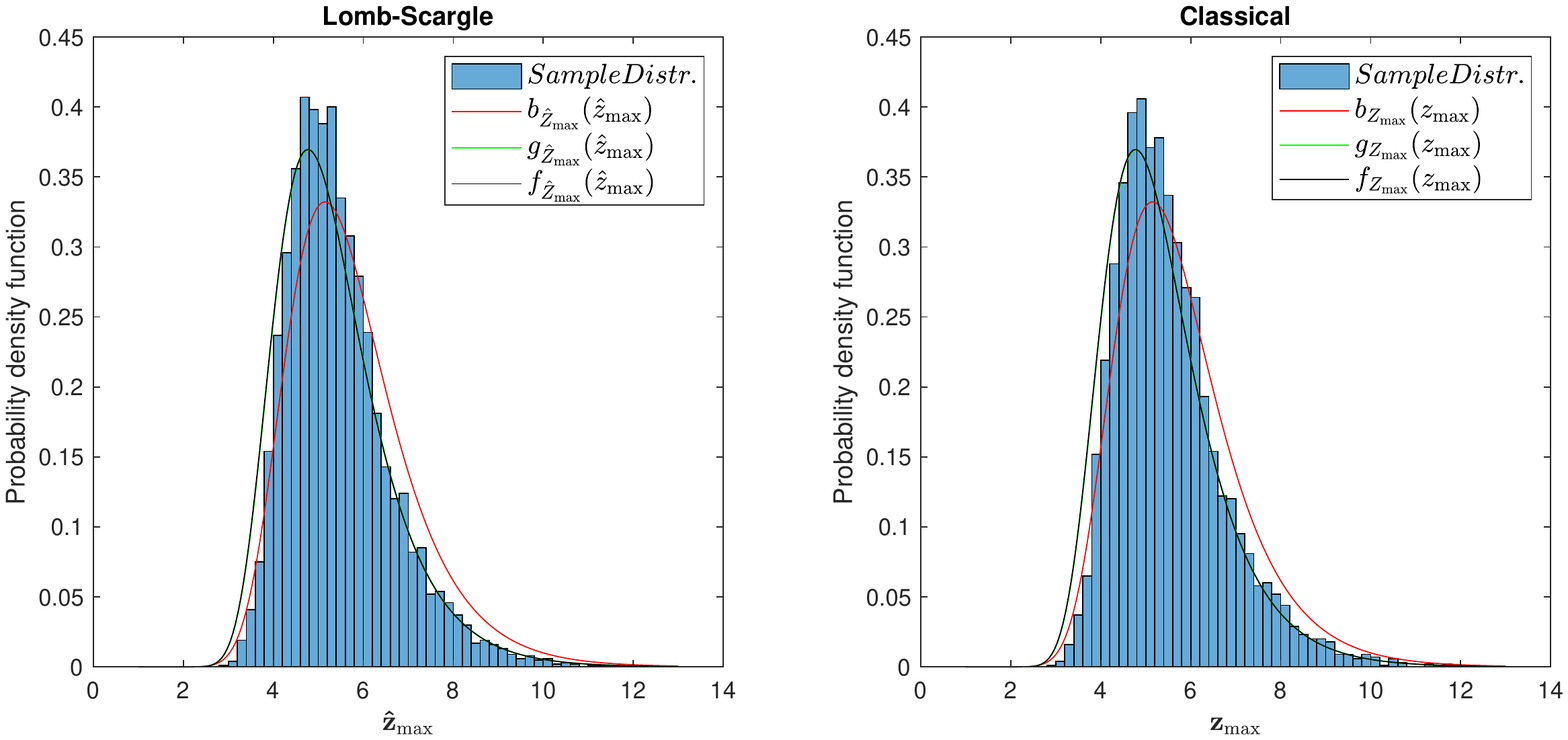}}
\vskip -1cm
        \caption{Histogram of the greatest peak for both the LSP and the CP obtained from $5000$ numerical simulations of the experiment in Fig.~\ref{fig:fig_vityazev2}
when the time series contains only a zero-mean, unit-variance white noise. The PDF $\bbf_{\hat{\Zf}_{\max}}(\hat{\zf}_{\max})$ derived from the extreme value theory by \citet{bal08}, the PDF 
$\gf_{\hat{\Zf}_{\max}}(\hat{\zf}_{\max})$ based on the exponential distribution given by Eq.~\eqref{eq:gpdf} and the proposed PDF $\ff_{\Zf_{\max}}(\zf_{\max})$ given by Eq.~\eqref{eq:gbpdf} are also plotted.}
        \label{fig:fig_SPFA2}
        \resizebox{\hsize}{!}{\includegraphics{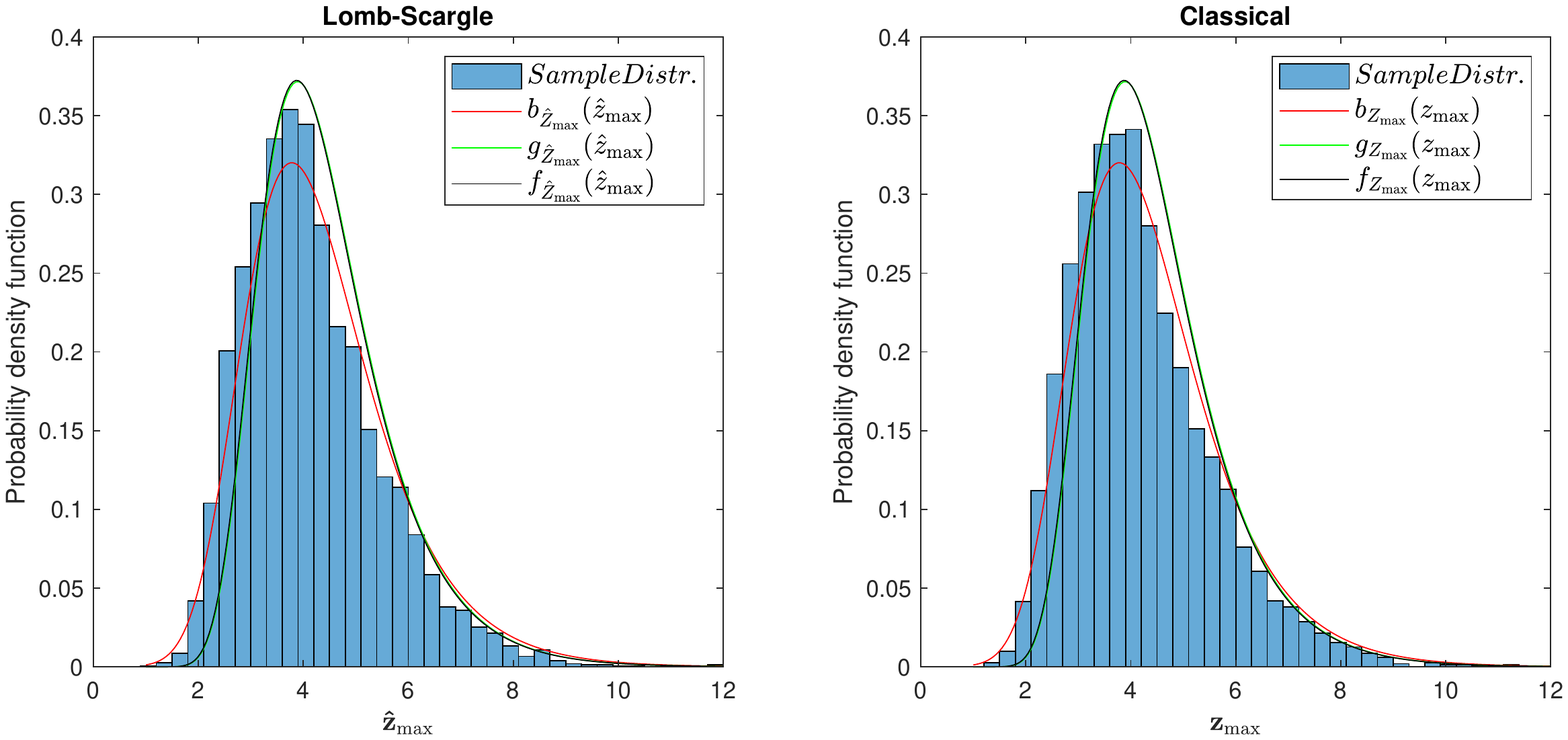}}
\vskip -1cm
        \caption{Histogram of the greatest peak for both the LSP and the CP obtained from $5000$ numerical simulations of the experiment in Fig.~\ref{fig:fig_sin}
when the time series contains only a zero-mean, unit-variance white noise. The PDF $\bbf_{\hat{\Zf}_{\max}}(\hat{\zf}_{\max})$ derived from the extreme value theory by \citet{bal08}, the PDF 
$\gf_{\hat{\Zf}_{\max}}(\hat{\zf}_{\max})$ based on the exponential distribution given by Eq.~\eqref{eq:gpdf} and the proposed PDF $\ff_{\Zf_{\max}}(\zf_{\max})$ given by Eq.~\eqref{eq:gbpdf} are also plotted.}
        \label{fig:fig_SPFA3}
\end{figure}

In order to test the reliability of the SPFA when estimated by means of analytical methods, Figs.~\ref{fig:fig_SPFA1}-\ref{fig:fig_SPFA3} show the histogram of the values $\hat{\zf}_{\max}$  and $\zf_{\max}$ of the highest peak of, respectively, the LSP and the CP obtained from $5000$ numerical simulations of a zero-mean, unit-variance white-noise process with the time sampling and the frequency pattern identical, respectively, to the experiments used for
Figs.~\ref{fig:fig_vityazev1}, \ref{fig:fig_vityazev2} and \ref{fig:fig_sin}. Three methods has been used to estimate the corresponding analytical PDF. In particular, the PDF $\gf_{\hat{\Zf}_{\max}}(\hat{\zf}_{\max})$ corresponding to the CDF $G^{N^*}_{\hat{Z}_{\max}}(\hat{\zf}_{\max})$ given in Eq.~\eqref{eq:pfaGN}
\begin{equation} \label{eq:gpdf}
\gf_{\hat{\Zf}_{\max}}(\hat{\zf}_{\max}) = N^* {\rm e}^{-\hat{\zf}_{\max}} \left(1 - {\rm e}^{-\hat{\zf}_{\max}} \right)^{N^*-1},
\end{equation}
the PDF $\ff_{{\Zf}_{\max}}({\zf}_{\max})$ corresponding to the CDF $F_{{\Zf}_{\max}}^{N^*}({\zf}_{\max})$ in Eq.~\eqref{eq:Fz}
\begin{equation} \label{eq:gbpdf}
\ff_{{\Zf}_{\max}}({\zf}_{\max}) = N^* \bar{f}_{{\Zf}_{\max}}({\zf}_{\max}) \bar{F}_{{\Zf}_{\max}}^{N^*-1}({\zf}_{\max}),
\end{equation}
where $N^*$ is computed
via the method presented in appendix~\ref{sec:appB}, and the PDF $\bbf_{\hat{\Zf}_{\max}}(\hat{\zf}_{\max})$  derived from the extreme value theory by \citet{bal08}.

From these figure it is possible to derive the following. The PDFs~\eqref{eq:gpdf} and \eqref{eq:gbpdf} are almost identical.  As mentioned above this is the consequence of the small correlation among the coefficients $a_{\nu}$ and $b_{\nu}$ for the vast majority of the frequencies.
Secondly, not unexpectedly, the sample distribution of the maxima of the LSP and the CP are very similar. Finally in the first two experiments, contrary to what has been found by \citet{suv15} in another set of experiments, the PDF computed
by means of Eq.~\ref{eq:pfaG1N} overperforms the approach by \citet{bal08}, whereas  the reverse holds for the third experiment. In this last case, however, the better approximation
concerns only the left tail, whereas the right tail (the most important for the estimation of the SPFA) are similar. This is the consequence of the fact that the approach based on the
extreme value theory provides only an upper limit to the SPFA and it requires that the LSP is free from strong aliasing effects. This condition is not satisfied in the present experiments.

\section{Application to an astronomical time series} \label{sec:data}

\begin{landscape}
\begin{figure}
        \resizebox{\hsize}{!}{\includegraphics{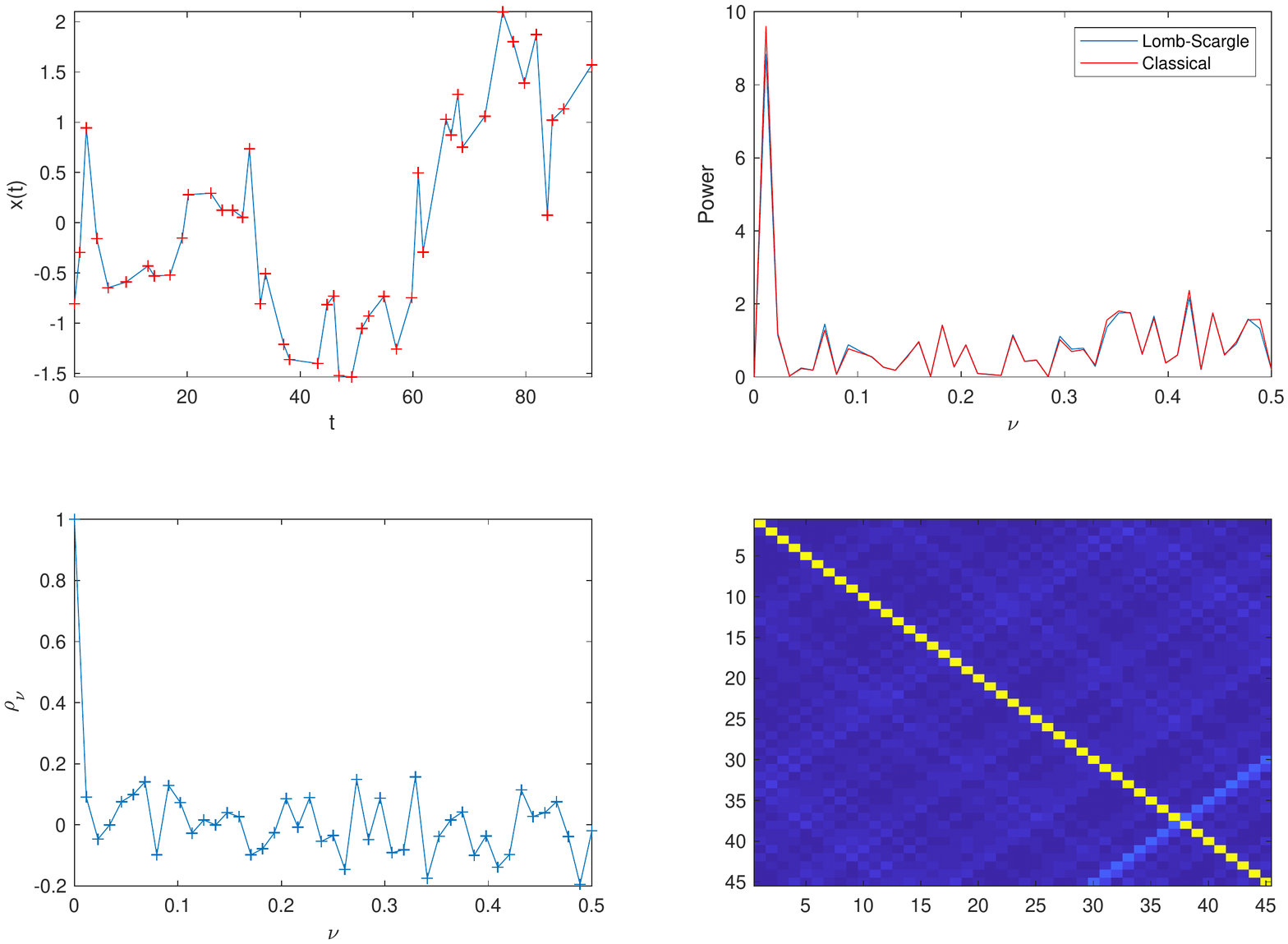}}
        \caption{Top-left panel: experimental time series  (standardized to zero mean and unit variance) containing $45$ unevenly-spaced data from \citet{big01}. Top-right panel: corresponding classical and Lomb-Scargle spectrograms 
computed on $45$ evenly spaced frequencies
in the range $[0, 0.5]$ in units of the Nyquist frequency. Bottom-left panel: correlation coefficients $\rho_{\nu}$ between the coefficients $a_{\nu}$ and $b_{\nu}$ for the frequencies used in the periodogram. Bottom-right panel: correlation matrix $\bf{\Sigma}$
for the frequencies used in the periodogram. This matrix is dominated by the diagonal entries and results of full rank. This indicates that all the frequencies are mutually uncorrelated.}
        \label{fig:fig_data}
\end{figure}
\end{landscape}

As a comparison of the performance of two methods in real experimental situations, the proposed procedures are applied to the time series (standardized to zero-mean and unit- variance) in the top-left panel of Fig.~\ref{fig:fig_data}. 
This time series is characterized by a rather irregular sampling. It was obtained with the VLA array \citep{big01} and consists
of polarisation position angle measurements at an observing frequency of 15~GHz for one of the images of the double gravitational lens system B0218+357. The corresponding CP and 
LSP are shown in the top-right panel of Fig.~\ref{fig:fig_data}.
They have been computed on a set of $45$ frequencies in the range $[0, 0.5]$ in unit of the Nyquist frequency (this last estimated on the basis of the shortest time distance between two observations).
The similarity of the two periodograms is apparent. As explained above, this is due to the fact that, as it is visible in the bottom-left panel of Fig.~\ref{fig:fig_data}, $|\rho_{\nu}|$  is small for all the frequencies. A prominent peak of amplitude 
$\approx 9.6$ appears at $\nu=1.37 \times 10^{-2}$ in both periodograms. The correlation matrix $\bf{\Sigma}$ shown in the bottom-right panel of the same figure results 
of full rank, hence $N^*=45$ and the SPFAs as computed by means of Eqs.~\eqref{eq:pfaG1N} for both the CP and the LSP  are $6.5 \times 10^{-3}$ and $2.8 \times 10^{-3}$, respectively. As expected, they are of the same order of magnitude.

\section{Conclusions} \label{sec:conclusions}

The Lomb-Scargle periodogram (LSP) has achieved great popularity thanks to its alleged capability of providing a statistical solution to the problem of the detection of 
signals embedded in noise when the sampling is irregular. Actually, this approach is based on the arbitrary procedure of forcing the entries of the periodogram to share the same exponential distribution. The appealing benefit is that
it is possible to access to a number of statistical tools for assessing the reliability of a detection which are not available otherwise.
However, in this way some theoretical difficulties are introduced which make its use difficult in certain steps of the detection procedure (e.g. the computation of the spectral windows), in the development of specific algorithms to compute the periodogram itself and the fact that, as it has been shown here, under certain conditions some statistically significant detections can be lost with respect to the classical periodogram (CP).
For these reasons, and because of the fact that typically the LSP and the CP provide almost identical results, there is no reason why the former should be considered superior to the latter.

\section*{Acknowledgements}






\appendix

\section{Why the Lomb-Scargle periodogram is not based on a correct least-squares model} \label{sec:appC}

Given an experimental uneven time series $\xb=[x_{t_0}, x_{t_1}, \ldots, x_{t_{M-1}}]$, testing the presence of an unknown signal 
$\ssb = [s_{t_0}, s_{t_1}, \ldots, s_{t_{M-1}}]$ is based on the following assumptions:
\begin{enumerate}
\item The signal is embedded in an additive noise $\nb$, i.e. the time series $\xb$ is given by $\xb = \ssb + \nb$. Without loss of generality, it is assumed that ${\rm E}[\nb] = 0$, where ${\rm E}[.]$ denotes the expectation operator; \\
\item The noise $\nb$ is the realization of a stationary stochastic process here assumed to be of white type.
\end{enumerate}   
An approach based on the least-squares modeling can be adopted. In this case, if signal $\ssb$ is unknown, it has to be assumed to have the form
\begin{equation} \label{eq:model}
s(t_j) = \sum_{k=0}^{N-1} \left[a_{\nu_k} \cos{(2 \pi \nu_k t_j)} + b_{\nu_k} \sin{(2 \pi \nu_k t_j}) \right],
\end{equation}
with $j=0, 1, \ldots, M-1$. Of course, part of the coefficients $\{a_{\nu_k}\}$ and $\{b_{\nu_k}\}$ could actually be equal to zero. For example, this happens with periodic signals which are not pure sinusoids.
However, in general this piece of information is not available in advance.
As a consequence, assuming that the frequencies $\{\nu_k\}$ are fixed, the correct least-squares model for the estimation the coefficients $\{a_{\nu_k}\}$ and $\{b_{\nu_k}\}$ is  
\footnote{We recall that the function 
``$ \underset{x}{\arg\min}[ H(x)]$'' provides the value of $x$ for which the function $H(x)$ has the smallest value.}
\begin{multline} \label{eq:corr}
[\{\tilde{a}_{\nu_k}\}, \{\tilde{b}_{\nu_k}\}] = \\
\underset{\{a_{\nu_k}\}, \{b_{\nu_k}\}}{\arg\min} \left\{ \sum_{j=0}^{M-1} \left[x(t_j) - \sum_{k=0}^{N-1} \left[a_{\nu_k} \cos{(2 \pi \nu_k t_j)} - b_{\nu_k} \sin{(2 \pi \nu_k t_j})\right] \right]^2\right\}.
\end{multline}
This means that all the coefficients $\{a_{\nu_k}\}$ and $\{b_{\nu_k}\}$ have to be simultaneously estimated.  A detection can be claimed if any of these coefficients results statistically different from zero.

This model is different from that implicitly assumed by the LSP, i.e.
 \begin{multline} \label{eq:lsp}
[\bar{a}_{\nu_k},\bar{b}_{\nu_k}] = \\
\underset{a_{\nu_k}, b_{\nu_k}}{\arg\min} \left\{ \sum_{j=0}^{M-1} \left[x(t_j) - a_{\nu_k} \cos{(2 \pi \nu_k t_j)} - b_{\nu_k} \sin{(2 \pi \nu_k t_j})\right]^2\right\}.
\end{multline}
It corresponds to assume that $\xb$ contains the contribution of one (and only one) pure sinusoidal signal $\ssb$ of fixed frequency not due to the noise. 
Here, the point is that,  for a given frequency $\nu_k$, the coefficients  $\bar{a}_{\nu_k}$ and $\bar{b}_{\nu_k}$ are identical to the corresponding $\tilde{a}_{\nu_k}$ and $\tilde{b}_{\nu_k}$ iff the set of sinusoidal functions constitutes an orthonormal basis \citep[e.g. see ][p. 450]{ham73}. This does not happen when the sampling is irregular. Hence, in most practical applications, the LSP is not based on the correct least-squares model.

\section{Number of uncorrelated frequencies in a periodogram} \label{sec:appB}

In the case of an unequally spaced time series $\{ x_{t_i} \}_{i=0}^{M-1}$ given by the realization of a discrete zero-mean, unit-variance, Gaussian white-noise process, the correlation $\rho_{12}$ between the values of the periodogram
in Eq.~\eqref{eq:pf} at two frequencies $\nu_1$ and $\nu_2$ is given by \footnote{They assumption of $\sigma_n=1$ for the white-noise process $\{ x_{t_i} \}$ is not a limitation since $\rho_{12}$ is independent of this parameter.},
\begin{equation} \label{eq:corr1}
\rho_{12} = \frac{{\rm E}\left[(a_1^2 + b_1^2 - \mu_{a_1^2} -\mu_{b_1^2}) (a_2^2 + b_2^2 - \mu_{a_2^2} -\mu_{b_2^2})\right]}{\sqrt{{\rm E}\left[(a_1^2 + b_1^2 - \mu_{a_1^2} -\mu_{b_1^2})^2\right]}
\sqrt{{\rm E}\left[(a_2^2 + b_2^2 - \mu_{a_2^2} -\mu_{b_2^2})^2\right]}},
\end{equation}
with $\mu_{a_i^2}$ and $\mu_{b_i^2}$ the expected values of $a_i^2$ and $b_i^2$,
\begin{align}
 \mu_{a_j^2} &= \sum_{i=0}^{M-1} \cos^2{(2 \pi \nu_j t_i)}, \\
 \mu_{b_j^2} &= \sum_{i=0}^{M-1} \sin^2{(2 \pi \nu_j t_i)}.
\end{align}
When Eq.~\ref{eq:corr1} is expanded, terms as
\begin{multline}
{\rm E}[a^2_1 a^2_2] = \sum_{i,j,l,m=0}^{M-1} {\rm E}[x_i x_j x_l x_m \cos{(2 \pi \nu_1 t_i)}  \cos{(2 \pi \nu_1 t_j)} \\ \times \cos{(2 \pi \nu_2 t_l)} \cos{(2 \pi \nu_2 t_m)}],
\end{multline}  
and similar ones must be evaluated considering four different situations, i.e. $i=j$ and $l=m$, $i=l$ and $j=m$, $i=m$ and $j=l$, and $i=j=l=m$. Keeping present this fact, after some algebra it is possible to show that
\begin{equation} \label{eq:rho}
\rho_{12} = \frac{\varrho^2_{a_1 a_2} \varrho^2_{a_1 b_2} \varrho^2_{b_1 a_2} \varrho^2_{b_1 b_2}}{\sqrt{\varrho^2_{a_1 a_1}+\varrho^2_{b_1 b_1}+2\varrho^2_{a_1 b_1}}{\sqrt{\varrho^2_{a_2 a_2}+\varrho^2_{b_2 b_2}+2\varrho^2_{a_2 b_2}}}},
\end{equation}
with
\begin{align}
\varrho_{a_l a_n} &= \sum_{i=0}^{M-1} \cos{(2 \pi \nu_l t_i)}  \cos{(2 \pi \nu_n t_i)}, \\
\varrho_{a_l b_n} &= \sum_{i=0}^{M-1} \cos{(2 \pi \nu_l t_i)}  \sin{(2 \pi \nu_n t_i)}, \\
\varrho_{b_l a_n} &= \sum_{i=0}^{M-1} \sin{(2 \pi \nu_l t_i)}  \cos{(2 \pi \nu_n t_i)}, \\
\varrho_{b_l b_n} &= \sum_{i=0}^{M-1} \sin{(2 \pi \nu_l t_i)}  \sin{(2 \pi \nu_n t_i)}.
\end{align}
If the periodogram is compute on a set of $N$ frequencies, from Eq.~\eqref{eq:corr1} it is possible to construct the correlation matrix
\begin{equation} \label{eq:matrix}
{\bf \Sigma} = 
\left( \begin{array}{ccccc}
\varrho_{11} &\varrho_{12}  &\cdots & \varrho_{1 (N-1)} & \varrho_{1N} \\
\varrho_{21} &\varrho_{22} &\cdots & \varrho_{2 (N-1)} & \varrho_{2N} \\
\vdots &\vdots & \ddots & \vdots & \vdots \\
\varrho_{(N-1) 1} & \varrho_{(N-1) 2} & \cdots & \varrho_{(N-1) (N-1)} & \varrho_{(N-1) N} \\
\varrho_{N1} & \varrho_{N2} & \cdots &  \varrho_{N (N-1)} & \varrho_{NN}
\end{array} \right),
\end{equation}
where $\varrho_{i i}=1$ and $\varrho_{ij} = \varrho_{ji}$. Now, the correlation matrix $\bf \Sigma$ of a set of $N$ uncorrelated random quantities is of full rank \footnote{The rank of a generic matrix  $\bf A$, ${\rm rank}[\bf A]$, is defined as 
the number of its linearly independent column vectors.} (indeed, it is the identity matrix), i.e. ${\rm rank}[{\bf \Sigma}]=N$. Hence, if ${\rm rank}[{\bf \Sigma}] = N^*$, $N^*$ provides the number of uncorrelated frequencies.
Actually, strictly speaking, $N^*$ computed in this way is not equivalent to the number of independent frequencies (uncorrelatedness does not imply independency). Indeed, in situations where $N > M$ it could happen that $N^* > M$, but
this does not mean that the number of independent frequencies is $N^*$. This is because the number of independent frequencies can be as large as the number of degrees of freedom of the system 
(in this case the number of data) at most. Therefore, when $N^* > M$ the number
of independent frequencies has to be assumed equal to $M$. In other words, $M$ has to be intended as an upper limit to the number of independent frequencies.

Typically, the numerical computation of ${\rm rank}[{\bf \Sigma}]$ is based on the number of singular values \footnote{A generic $N \times N$ real matrix $\bf A$ can always be factorized into the form ${\bf A}={\bf UDV}^T$ where $\bf U$ and $\bf V$ are $N \times N$ orthogonal matrices, whereas $\bf D$ is an $N \times N$ diagonal matrix whose diagonal elements $\delta_1 \ge \delta_2 \ge \ldots \ge \delta_N \ge 0$ are known as singular values \citep{bjo96}.} 
greater than a given tolerance ${\it tol}$. There are various choices to fix this quantity. An example is  ${\it tol} = N \epsilon ||{\bf \Sigma}||_2$, where $\epsilon$ is the unit roundoff error ($=2^{-52}$) and  $||{\bf \Sigma}||_2$ is given by the greatest singular values of ${\bf \Sigma}$ \citep[see pag. 268 in][]{wat10}. 

\section{PDF and CDF of the entries of the periodogram of a discrete, uneven, zero-mean, white-noise process.} \label{sec:appA}

In the case of a discrete, uneven, zero-mean, white-noise process the PDF of the periodogram $p_{\nu}$ at a specific frequency is given by the PDF $f_Z(z)$ of the random variable
$Z=X_1^2+X_2^2$ with $X_1$ and $X_2$ zero-mean, Gaussian random variables  with standard deviation $\sigma_1$ and $\sigma_2$, respectively, and correlation coefficient $\rho$. This PDF can be obtained
considering that $Z$ can be expressed as the sum of two independent, squared, zero-mean, Gaussian random variables $\bx_1 \sim  \mathcal{N}(0, \bbsigma^2_1)$ and $\bx_2 \sim  \mathcal{N}(0, \bbsigma^2_2)$ with
$\bbsigma^2_1$ and $\bbsigma^2_2$ the eigenvalues of the covariance matrix $\Sigma$ of $X_1$ and $X_2$,
\begin{equation} \label{eq:matrix}
{\bf \Sigma} = 
\left( \begin{array}{cc}
\sigma_1^2 &  \rho \sigma_1 \sigma_2 \\
 \rho \sigma_1 \sigma_2 & \sigma_2^2 \\
\end{array} \right).
\end{equation}
In formula,
\begin{align}
\bbsigma^2_1 & = \frac{1}{2} \left( \sigma_1^2 + \sigma_2^2 - \sqrt{\sigma_1^4 + \sigma_2^4+2 (2 \rho^2 -1) \sigma_1^2 \sigma_2^2} \right),  \label{eq:bsigma1} \\
\bbsigma^2_2 & = \frac{1}{2} \left( \sigma_1^2 + \sigma_2^2 + \sqrt{\sigma_1^4 + \sigma_2^4+2 (2 \rho^2 -1) \sigma_1^2 \sigma_2^2} \right).  \label{eq:bsigma2}
\end{align}

Since $\bx_i / \bbsigma_i \sim \mathcal{N}(0, 1)$, it happens that  $\bx_i^2 / \bbsigma_i^2 \sim \bbchi^2_1$ with $\bbchi^2_1$ the chi-squared distribution with one degree of freedom. In its turn, $ \bbchi^2_1 = \Gamma(1/2, 2)$ with
\begin{equation}  \label{eq:CDF}
\Gamma(a,b)=
\begin{cases}
\frac{b^a}{\Gamma(a)} \tau^{a-1} {\rm e}^{-b \tau} &  \text{if } \tau \ge 0, \\
0   &  \text{if }  \tau < 0,
\end{cases}
\end{equation}
the Gamma PDF.
Hence, $\bx_i^2 \sim \bbsigma_i^2 \Gamma (1/2,2) = \Gamma(1/2, 2 \bbsigma_i^2)$.
As a consequence, $f_Z(z)$ can be obtained by the convolution of two Gamma PDFs, $\Gamma(a_1, b_1)$ and $\Gamma(a_2,b_2)$,
\begin{equation}
f_Z(z) = \frac{b_1^{a_1} b_2^{a_2}}{\Gamma(a_1) \Gamma(a_2)} \int_0^{z} \exp{\left[-b_1 (z-y)-b_2 y \right]} (z-y)^{a_1-1} y^{a_2-1} dy,
\end{equation}
with $a_1=a_2=1/2$, $b_1=1/(2 \bbsigma^2_1)$ and $b_2=1/(2 \bbsigma^2_2)$.
The change of variable $y= z t$ provides
\begin{equation}
 f_Z(z) = \frac{b_1^{a_1} b_2^{a_2}  z^{a_1+a_2-1}}{\Gamma(a_1) \Gamma(a_2)} {\rm e}^{-b_1 z} \int_0^{1} \exp{\left[ (b_1 - b_2) z t \right]} (1-t)^{a_1-1} t^{a_2-1} dt,
\end{equation}
which can be written in the equivalent  form
\begin{equation}
f_Z(z) = \frac{b_1^{a_1} b_2^{a_2} z^{a_1+a_2-1}}{\Gamma(a_1 + a_2)} {\rm e}^{-b_1 z} {}_{\phantom{1}1}F_1\left(a_2,a_1+a_2; (b_1 - b_2) z\right),
\end{equation}
with ${}_{\phantom{1}1}F_1(.,.;.)$ the Kummer confluent hypergeometric function. Finally,  since
\begin{equation}
{}_{\phantom{1}1}F_1\left[\frac{1}{2},1; \gamma z \right]=\exp{\left[\frac{\gamma z}{2} \right]} I_0\left[\frac{\gamma z}{2} \right], 
\end{equation}
with $I_0(.)$ the modified Bessel function of the first kind and zero order, and taking into account that in the present case
\begin{equation}
\gamma = \frac{\bbsigma^2_2 -  \bbsigma^2_1}{2  \bbsigma^2_1  \bbsigma^2_2} ,
\end{equation}
$f_Z(z)$ takes the form
\begin{equation}  \label{eq:fzz}
f_Z(z)=
\begin{cases} 
\exp{\left[-\left(\frac{ \bbsigma^2_1+ \bbsigma^2_2}{4 \bbsigma^2_1  \bbsigma^2_2} \right) z \right]}   \frac{1}{2 \bbsigma_1  \bbsigma_2} I_0 \left[ \frac{ \bbsigma^2_2 -  \bbsigma^2_1}{4 \bbsigma^2_1  \bbsigma^2_2} z \right] &  \text{if } z \ge 0,  \\  
0  &  \text{if }  z < 0.
\end{cases}
\end{equation}

A useful approximation for $f_Z(z)$ can be obtained from the fact that for sufficiently great values of $y$, $I_0(y)$ can be well approximated by \citep{zha96}
\begin{equation} \label{eq:fza}
I_0(y) \approx \frac{{\rm e}^y}{\sqrt{2 \pi y}}.
\end{equation}
Now, in the case of strongly correlated Gaussian random variable ($|\rho| \approx 1$), it happens that $\bbsigma^2_2 \gg \bbsigma^2_1\approx 0$. Under this condition, it is possible to see that $f_Z(z)$, as given
by Eq.~\eqref{eq:fzz} with Eq.~\eqref{eq:fza}, can be approximated by
\begin{equation} 
f_Z(z) \approx \frac{1}{\bbsigma_2 \sqrt{2 \pi z}} {\rm e}^{-z/(2 \bbsigma_2^2)}.
\end{equation} 

The CDF $F_Z(z)$  can be obtained  by direct numerical integration of Eq.~(\ref{eq:fzz}) or by recognizing that, up to a constant  $1/(2 \bbsigma_1  \bbsigma_2)$, the required integral is of type 
$\int_0^{z} \exp{(-p u)} I_0(q u) du$ with solution given by  Eq.~(25) of the table of integrals in \cite{nut72},
\begin{multline}
\int_0^{z} \exp{(-p u)} I_0(q u) du  = \\ \frac{1}{s} \left[1 + \exp{(-p z)} I_0(q z) - 2 Q_1(v,w) \right], \quad p \neq q,
\end{multline}
with $Q_1(.,.)$ the Marcum Q-function, $s=\sqrt{p^2-q^2}$, $v=\sqrt{z(p-s)}$ and $w=\sqrt{z(p+s)}$. Here, $p=- (\bbsigma^2_1+ \bbsigma^2_2)/(4  \bbsigma^2_1  \bbsigma^2_2)$ and $q = (\bbsigma^2_2 -  \bbsigma^2_1)/(4  \bbsigma^2_1  \bbsigma^2_2)$.

\end{document}